\documentclass[a4paper,fleqn]{cas-dc} 

\usepackage[authoryear,longnamesfirst]{natbib}
\usepackage{pifont}
\usepackage[utf8]{inputenc}

\newcommand{\cmark}{\color{Green}\ding{51}\color{black}}%
\newcommand{\xmark}{\color{red}\ding{55}\color{black}}%

\newcolumntype{C}[1]{>{\raggedright\arraybackslash}p{#1}}
\def\tsc#1{\csdef{#1}{\textsc{\lowercase{#1}}\xspace}}
\tsc{WGM}
\tsc{QE}

\begin{document}
\let\WriteBookmarks\relax
\def\floatpagepagefraction{1}
\def\textpagefraction{.001}

\shorttitle{Inverse methods: How feasible are low-resolved modeling results?}    

\shortauthors{Martha Maria et~al.}

\title{Inverse methods: How feasible are spatially low-resolved capacity expansion modelling results when disaggregated at high spatial resolution?}
\tnotemark[1]

\tnotetext[1]{This document is the result of the research
	project funded by the {Helmholtz Association} under the program ``Energy System Design'' and under grant no. {VH-NG-135}.}

\author[kit,cor1]{Martha Maria Frysztacki}[orcid=0000-0002-0788-1328]
\fntext[cor1]{Corresponding author: \texttt{martha.frysztacki@kit.edu}}
\credit{Conceptualization, Methodology, Software, Formal Analysis, Data Curation, Writing - Original Draft, Writing - Review \& Editing, Visualization}

\author[kit]{Veit Hagenmeyer}
\credit{Writing - Review \& Editing, Funding Acquisition}

\author[tub]{Tom Brown}[orcid=0000-0001-5898-1911]
\credit{Conceptualization, Writing - Review \& Editing, Project Administration, Funding Acquistion}

\address[kit]{Institute for Automation and Applied Informatics, Karlsruhe Institute of Technology, 76344 Eggenstein-Leopoldshafen, Germany}

\address[tub]{Institute of Energy Technology, Technical University Berlin, Einsteinufer 25, 10587 Berlin}

\begin{abstract}
Spatially highly-resolved capacity expansion models are often simplified to a lower spatial resolution because they are computationally intensive. The simplification mixes sites with different renewable features while ignoring transmission lines that can cause congestion. As a consequence, the results may represent an infeasible system when the capacities are fed back at higher spatial detail. Thus far there has been no detailed investigation of how to disaggregate results and whether the spatially highly-resolved disaggregated model is feasible. This is challenging since there is no unique way to invert the clustering.

This article is split into two parts to tackle these challenges. First, methods to disaggregate spatially low-resolved results are presented: (a) an uniform distribution of regional results across its original highly-resolved regions, (b) a re-optimisation for each region separately, (c) an approach that minimises the ``excess electricity''. Second, the resulting highly-resolved models' feasibility is investigated by running an operational dispatch. While re-optimising yields the best results, the third inverse method provides comparable results for less computational effort. Feasibility-wise, the study design strengthens that modelling countries by single regions is insufficient. State-of-the-art reduced models with 100--200 regions for Europe still yield 3\%--7\% of load-shedding, depending on model resolution and inverse method.
\end{abstract}


\begin{highlights}
\item Methodology to disaggregate spatially low-resolved capacity expansion model results.
\item Feasibility analysis of the 100\% renewable, disaggregated, highly-resolved model.
\item One-node-per-country models of Europe are not fit for accurate investment decisions.
\item Results from 100--200 node models of Europe lead to more than 3\%--7\% of load-shedding.
\end{highlights}

\begin{keywords}
Electricity system optimisation \sep Renewable energy \sep Investment planning \sep Spatial clustering \sep Inverse methods \sep Disaggregation methods
\end{keywords}

\maketitle

\section{Introduction}\label{sec:introduction}
Capacity expansion models are computer programs for the dimensioning of energy generators, storage or the expansion of transmission lines, typically utilised to simulate ambitious climate change mitigation or carbon-dioxide reduction targets. Such models have gained recognition over the past years as their results are often taken as a reference to formulate energy transition road maps.
They are used by non-governmental organisations, political institutions, transmission system operators and large companies.
Latest trends in model utilisation are analysed by~\citet{lopion2018}, who
outline that today's models must accurately portray renewable potentials and
thus capture the weather-driven variability of wind and solar photovoltaic
generation in order to provide reliable investment recommendations for
renewable generator, storage and transmission installations.

Recent literature has identified that a high spatial resolution is required for the modelling to produce an accurate representation of renewable generation.
\citet{Schlachtberger2017} found that modelling a large geographical area at the
scale of Europe allows the model to exploit very good continental renewable
potentials and strongly impacts the composition of the generation and storage
fleet of individual regions.~\citet{frew2016} demonstrated similar findings for
the United States.
For models at continental scope,~\citet{Aryanpur2021} stress the incompleteness
of single region optimisation models that do not capture sufficient spatial
detail. They demonstrate that a high spatial granularity is particularly
relevant for models that include heterogeneous regions (such as regions with a
high share of renewable generation), or regions with higher variability in
energy demands. In such modelling scenarios, results significantly change when
the spatial resolution of the model is varied.
\citet{martinzegordon2021} explain the different modelling results by the
ability of a spatially detailed model resolution to detect bottlenecks in the
transmission grid and, therefore, to adequately assess renewable potentials
based on local weather conditions and to identify regional variations in
electricity demand.
The granularity of this data-driven information significantly improves the design and composition of the electricity mix and the routing of new grid infrastructure predicted by the model.
Similar results have been found by~\citet{miranda2019}, who show that a detailed
transmission infrastructure significantly affects capacity deployments and
electricity prices.
On the resource side,~\citet{frew2016} reveal significant differences in wind
and solar development when modelling renewable generation sites independently
for optimal site diversity, compared to assuming an aggregated buildout across
all sites uniformly as is done in many grid integration studies.
\citet{frysztacki2021} disentangle the effects of sourcing renewable generation
versus routing the electricity to locations of high demand using large
transmission networks, revealing that routing dominates the system effects and
forces the model to build renewable assets closer to demand centres at
potentially worse capacity factors when the model has a high spatial
resolution.
These findings are not unique to the electricity sector; they can also be
observed when modelling the heat decarbonisation \cite{jalilvega2018}.

The downside of higher resolution is that processing large data volumes inevitably results in a computational burden that arises from solving the associated mathematical formulation in the model.
In order for the spatially aggregated models to better represent the highly-resolved system, more effective clustering methods have been developed.
Methods developed by~\citet{scaramuzzino2019} and \citet{Siala2019} show that
aggregating based on political zones and borders is not suited to accurately
portray the electricity system because there is per se no correlation between
the distribution of solar radiation, wind speed, electrical load on the one
hand and the administrative divisions on the other hand.
Instead, they suggest to aggregate regions based on their similarities in renewable feed-in or demand-patterns, such as load density distribution and solar and wind potentials.
Another approach by~\citet{BIENER2020106349} suggests to cluster regions with
high electrical connectivity to minimise load flow deviations after the
aggregation.
A recent survey by~\citet{Frysztacki2022} improves previous methods and suggests
which of them is most suitable for which modelling scenario.
The overall consensus of most studies is to model the European electricity system that contains more than $5000$ electrical substations at and above $220$ kV at a spatial model resolution of at least $100\text{--}200$ regions, depending on the model configuration.

Previous literature did not provide answers if the spatial resolution of the models impacts the optimality of the solution, or whether these spatially simplified modelling results are feasible with respect to the original, spatially highly-resolved model.
Moreover, as of now, there exist no or only limited approaches in previous
research to disaggregate the spatially simplified modelling results back at its
original high resolution. There exist only few publications to the authors
knowledge, such as from~\cite{UlfMueller2019} who disaggregate simplified
modelling results at higher spatial detail. In their approach only the power
dispatch of spatially low-resolved model was disaggregated to a spatially
higher resolved model, investment variables were not considered. Beyond that,
the study did not analyse the overall feasibility of the resulting
disaggregated model at a higher spatial resolution and focuses only on a German
model. Neither the assumptions on the disaggregation, or the resulting
highly-resolved systems were analysed in detail with respect to their
plausibility.
Another disaggregation method is presented by~\citet{Reinert2020}, who propose
an iterative disaggregation process and allow transmission grid expansion in
the final iteration, so that solving the highly-resolved model becomes
feasible. It remains unclear how feasible the disaggregated model is when
omitting transmission grid expansion in the final iteration.
Finally,~\citet{Grochowicz2022} propose three disaggregation methods. At highest
resolution, they generate results for a model where every country is
represented as a single region, which is already below the sufficient
resolution suggested by prior research. Moreover, the proposed disaggregation
follows a pre-defined iterative approach that guarantees that the resulting
highly-resolved one-region-per-country model is feasible. Therefore, they do
not provide an analysis on whether spatially low-resolved capacity planning
results can represent a feasible solution to a spatially more complex system.

This contribution focuses on the unresolved inverse problem of mapping spatially low-resolved optimisation variables at a higher resolution. The first comparison of different disaggregation methods in a European continent-scale model is presented and new algorithms for disaggregation options are provided in  \ref{sec:methods}.
Spatial decomposition methods such as optimality cut or feasibility cut were specifically not applied, because this approach would aim to solve the fully resolved model, and provide no insight if the spatially aggregated model results are feasible with respect to the higher dimensioned model formulation.

First, the capacity expansion model used for this study is presented, followed by a detailed description of three new methods to disaggregate the resulting coarse model variables at high resolution in Section \ref{sec:methods}.
In Section \ref{sec:results} a feasibility analysis is carried out where it is tested if the spatially low-resolved capacity planning results are capable to meet electricity demand at all places at all times when disaggregated at a spatially highly-resolved model. The feasibility of the resulting system is not intuitive, because the spatially low-resolved models omit transmission constraints that can result in congestion and, at the same time, smooth variable renewable resources by aggregating potentially heterogeneous resource sites.
The novelty of this analysis is that none of the very limited disaggregation methods from the literature was previously tested with respect to model feasibility. Therefore, the error in investment decisions made by the low-resolved models compared to a spatially highly-resolved model is quantified for the first time.
Conclusions are drawn in Section \ref{sec:conclusions}.
Finally, the limitations of this study are discussed in Section \ref{sec:limitations}.

\section{Data and Methods}\label{sec:methods}
In this section the underlying data is presented. The overall modelling process in described in Section \ref{sec:modeloverview}, where the novelty of this study is highlighted. A selection of the most important data and methods of the model employed for this study are presented in Section \ref{sec:modelanddata}. The three proposed inverse methods on how spatially low-resolved capacity expansion model results can be disaggregated back at higher spatial detail are presented in Sections \ref{sec:uniform}--\ref{sec:minexcess} and are summarised in Table
\ref{tab:methods}. The treatment of inter-cluster powerflows is discussed in Section \ref{sec:powerflows}. The study design and evaluation of results are presented in Section \ref{sec:studydesign}.

\subsection{Modelling Overview} \label{sec:modeloverview}

\begin{figure}
	\centering
	\includegraphics[width=.45\textwidth]{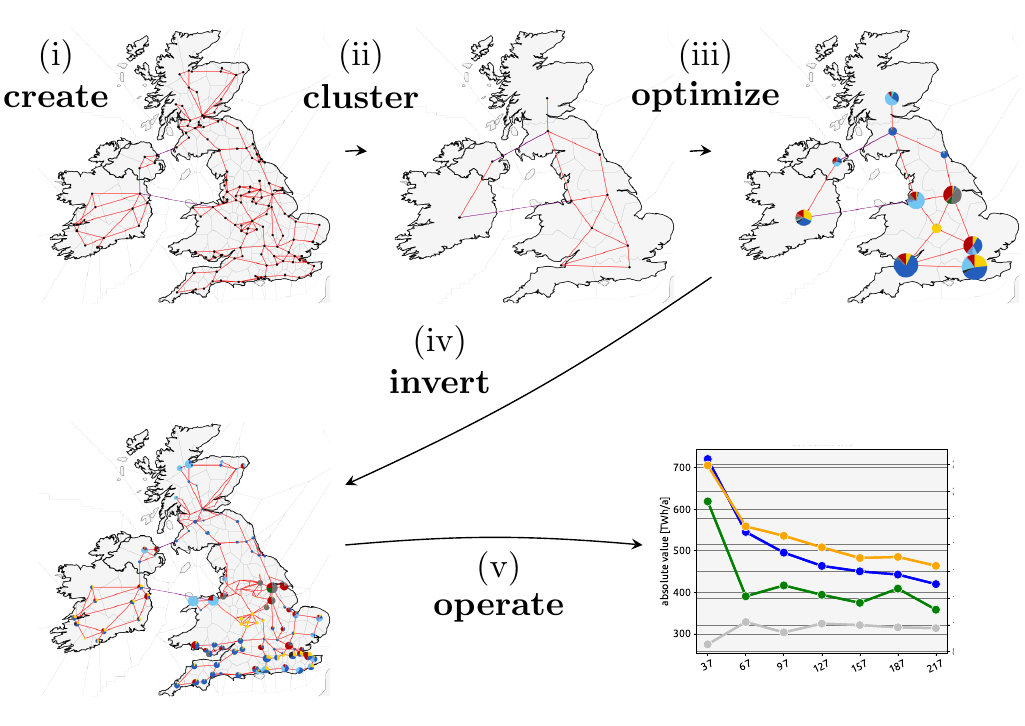}
	\caption{Illustration of the typical model process: (i) Data collection and model creation, (ii) spatially clustering the network down to a smaller number of nodes to obtain a computationally tractable model, (iii) optimise the model, i.e.~find the cost-optimal configuration of generators, storage units and power flows. Normally, the modelling task is completed at this stage. In this work, two additional steps are proposed: (iv) disaggregating the spatially reduced model results into higher spatial resolution and (v) an operational optimisation is conducted, testing the resulting network model for feasibility.}\label{fig:infographic}
\end{figure}

Figure \ref{fig:infographic} displays the overall approach of electricity system modelling. It typically executes in the following order: (i) Creating the model. This includes collecting data of the system to be analysed, for example the network topology of the transmission system, capacities of generators that are to be included in the model, land-use constraints, time-series of electricity demand, wind speeds, solar radiation, etc. and assigning the data to the correct locations. (ii) Clustering the spatially highly-resolved network down to a smaller approximation to gain computational advantages. (iii) Formulating a set of mathematical equations associated with the problem and solving it.

Here, the additional fourth and fifth steps are introduced to this queue: (iv) Disaggregating the low-resolved results (i.e.~the resulting renewable capacities $G_{v,s}$) back at high spatial resolution.
As the clustering $f: \mathbb{R}^m \rightarrow \mathbb{R}^n$
reduces the (spatial) dimension of the data ($n < m$, in many cases even $n \ll m$), the mapping is surjective but not injective, hence not bijective. Therefore, finding an inverse that maps the results back at high dimension $ f^{-1}: \mathbb{R}^n \rightarrow \mathbb{R}^m $
is a challenging task and the inverse is not unique. Therefore three different approaches to tackle the disaggregation of the generation and storage capacities for each technology from the low-resolved capacity expansion model to the highly-resolved operational model are presented and adequate inverse methods are suggested. The proposed methods are summarised in 
Table \ref{tab:methods} and are explained in detail in the following three sections.
(v) Running an optimisation with fixed capacity that is derived from step (iii), applying the disaggregation from (iv). (v) is also referred to as operational optimisation. It allows to gain insights into the dynamics of the electricity system, particularly to analyse its feasibility.

\subsection{Model and Input Data} \label{sec:modelanddata}
For this study, the openly available European Electricity System Model at
transmission substation level, PyPSA-Eur is employed. It is used to model a
future fully renewable electricity system that consists of today`s transmission
grid, and can build solar pv, onshore wind, offshore wind, hydrogen storage and
batteries to cover European electricity demands. The model is described in
detail in its original publication by~\citet{PyPSA-Eur}, therefore only its
major functionality are summarised here. The mathematical model formulation and
solving is based on the python package PyPSA, originally developed
by~\citet{PyPSA}. This section focuses on features that are relevant for the
specific use-case for this study.

The full model covers 33 European countries with a full spatial resolution of
$5323$ nodes ($3609$ substations), $6640$ transmission HVAC and
HVDC lines at and above $220$kV. The transmission grid is assumed as it
was installed by 2020, but includes planned over-sea HVDC lines that strengthen
the connection between continental Europe and the British Isles. Electricity
demand is embedded from the~\citet{OPSD} project. Historical weather data to
account for the variability of renewable resources is openly provided by
the~\citet{ERA5} in the ``ERA5 Reanalysis'' dataset, and by~\citet{SARAH2} in the
second edition of the ``Surface Radiation Data Set (SARAH-2)''. This raw data
contains for example solar radiation, wind-speeds or temperature, and is
processed using the open source software ``atlite'', developed
by~\citet{atlite}, to translate it into capacity factors for the modelling.

The main objective of the capacity expansion model is to minimise the annual
system costs that consist of the sum of all investments in new capacity
$G_{v,s}$ at each node $v$ of every technology $s$, as well
as the variable costs related to the dispatch of the generators $g_{v,s,t}$ at
each time $t$. The weight $w_t$ relates to the duration of
dispatch and is fixed to $2$ in our application, i.e.~$w_t \equiv 2$ to
balance computational model fidelity against accuracy of the modelling results
[see~\cite{SCHLACHTBERGER2018}]. Mathematically, this can be formulated as
\begin{align}
	\label{eq:objectivefunc}
	&  \min_{{G_{v,s},\atop  g_{v,s,t}},\atop f_{(v,w),t}} \Bigl[ \sum_{v\in\mathcal{V}} \sum_{s\in\mathcal{S}} \Bigl( c_{v,s} G_{v,s} + \sum_{t\in\mathcal{T}} w_t o_{v,s}g_{v,s,t} \Bigr) \Bigr] \,,
\end{align}
with additional constraints to guarantee network security, to cover demand at
all times and places, to account for the system to be physically plausible
including Kirchhoff's circuit laws and upper and lower bounds for generator
dispatch. Such model is called ``capacity expansion model'', because the
variables that are subject to the optimisation represent the dimensioning of
renewable generators or storage units, $G_{v,s}$. Furthermore, generator
dispatch and energy storage behaviour as well as electricity power flows are
also subject to the optimisation, constrained by the size of the respective
unit. The optimisation is solved using an interior point method to find the
minimum using the~\citet{gurobi} python interface.

All variables appearing in this article are explained in the Glossary, Section \ref{sec:glossary}, Table \ref{tab:glossary}. Cost assumptions are based on suggestions by the Danish Energy Agency
\cite{dea2019} for wind technologies,~\citet{schroeder2013} in case of open
cycle gas turbines, pumped hydro storage, hydro,
run-of-river,~\citet{budischak2013} for storage technologies and~\citet{etip} for
solar. For details, see \ref{appendix:costs}.

To enable computational feasibility, the whole model with more than $5000$
nodes must be reduced to a computationally tractable size by spatially
clustering the nodes. Then, the reduced optimisation problem
(\ref{eq:objectivefunc}), now with a smaller set of nodes $\mathcal{V}$, can be
solved. There exist many approaches to spatially reduce the size of such model.
Latest research from~\cite{BIENER2020106349,Frysztacki2022},
motivated by insights from~\cite{Siala2019}, have demonstrated that the best
suited clustering method for capacity expansion models with highly renewable
scenarios are of hierarchical nature. Therefore, for this study, a hierarchical
agglomerative clustering is chosen for the spatial scale. It is a bottom-up
approach, where each node is treated as a singleton cluster. Then, in every
iteration, two clusters that have the highest similarity and that are connected
by a transmission line are aggregated. The same similarity measure as
previously analysed is used. It is defined such that the aggregated nodes have
the most similar renewable time-series $\bar{g}_{v,s,t}$ throughout the whole year,
i.e.~\begin{equation}
	\label{eq:time}
	\bar{g}_{{v\in \mathcal{V},\
			s\in\{\mathrm{solar},\,\mathrm{wind}\},\
			t\in\mathcal{T}}} \in [0,1]^{2\mathcal{T}}\, .
\end{equation}

Note that such spatially aggregated models are not capable to account for all transmission lines that can be constraining in terms of the load-flow. For example, transmission lines within an aggregated region are ignored in the simulation. As a result, the optimal solution of a spatially clustered model can still be inaccurate compared to a highly-resolved solution. Eventually, this inaccuracy can lead to an infeasibility when coarse modelling results are implemented in a spatially more complex system. This inaccuracy is quantified in Section \ref{sec:results}. For the remaining part of this section, the focus lies on the three proposed disaggregation methods including the treatment of inter-cluster transmission flows.

In terms of temporal scale, every two consecutive hours are aggregated.
Temporal aggregation, such as spatial aggregation, can lead to a different type
of modelling error. Therefore a high temporal resolution is maintained. The
approach to aggregate only two consecutive hours has shown to yield good
results compared to hourly-resolution while reducing the model size by a factor
of $2$. A general overview of temporal aggregation and accompanying
errors are discussed by~\citet{KOTZUR2018474} or~\citet{Jacobson2022}.

\begin{table}
\caption{Summary of the proposed dis-aggregation methods.}\label{tab:methods}
\begin{tabular}{@{} lp{5.8cm}}
	\toprule
	\textbf{Short name} & \textbf{Method description} \\ 
	\midrule
	\ & \hspace*{-6.2em}Optimal capacities retrieved from an optimised low-resolu-\\
	\ & \hspace*{-6.2em}-tion model are distributed ... \vspace*{.5em} \\
	\textbf{uniform} & ... uniformly across all nodes within a cluster of the high-resolution network while accounting for land-use restrictions by imposing an upper bound. \\
	\textbf{re-optimize} & ... anew by re-optimizing capacities within each cluster with full formulation, while enforcing the same build-out capacity totals per technology as in the clustered model.\\
	\textbf{min excess} & ... according to a local optimisation which seeks to concentrate generation at nodes with higher demand and grid capacity and thus to minimise load-shedding.\\
	\bottomrule
\end{tabular}
\end{table}

\subsection{Uniform Distribution} \label{sec:uniform}

The first approach to disaggregate spatially low-resolved modelling results is
inspired by suggestions made by~\citet{UlfMueller2019}, where a similar
disaggregation was applied on the power dispatch $g_{c,s,t}$. Here, the focus
lies on the investment variables $G_{c,s}$ and the disaggregated dispatch is
determined only after running an operational problem (see
Section \ref{sec:modeloverview}).
The method is simple and computationally inexpensive. For each generation and storage technology, the capacities retrieved from the spatially low-resolved model is distributed within every cluster uniformly across all highly-resolved nodes:
\begin{equation}
	\label{eq:uniform}
	G_{c,s} \mapsto \frac{1}{\mathcal{V}_c}
	\left( 
	\begin{matrix}
		G_{c,s} \\ ...\\ G_{c,s}
	\end{matrix}
	\right) \in \mathbb{R}^{\mathcal{V}_c} \,.
\end{equation}
An additional constraint is formulated to account for land-use constraints. This upper limit is formally defined as
\begin{equation}
	G_{v,s}\leq G_{v,s}^{\mathrm{max}}\ \quad  \forall v\in \mathcal{V}_c \label{eq:uniform_constraints}
\end{equation}
and is enforced by selecting the generators where (\ref{eq:uniform_constraints}) is not satisfied and uniformly distributing the residual capacity over the remaining nodes within the cluster, i.e.~across the following set of generators:
\begin{equation}
	\label{eq:lanuseuniform}
	\{G_{v,s}: v\in \mathcal{V}_c \land G_{v,s} < G_{v,s}^{\mathrm{max}}\}\,.
\end{equation}
The last step is repeated until all nodes satisfy constraint (\ref{eq:uniform_constraints}).

\subsection{Regional Re-Optimisation} \label{sec:reopt}

The second considered method is inspired by suggestions made
by~\citet{Reinert2020}, who have proposed a similar, iterative disaggregation
method with adaptations to the transmission grid in the final iteration.
Further adaptations to the optimal solution or the original transmission grid
are not allowed, and the problem is not iterative. However, it may be
computationally challenging. For each cluster $\mathcal{V}_c\subset \mathcal{V}$ the re-optimisation
is conducted at high resolution using the original objective function
(\ref{eq:objectivefunc}) with all associated mathematical constraints.
Additionally, a set of constraints to incorporate the low-resolved modelling
results is imposed:
\begin{equation}
	\sum_{v\in \mathcal{V}_c} G_{v,s} = G_{c,s} \label{eq:reopt_eqconstraint} \quad \forall s\in \mathcal{S}\,.
\end{equation}
This set of constraints ensures that the amount of installed capacity is the same for every technology in every region.

Depending on the size of the cluster, the disaggregation may blow up the problem beyond the computational capacity of the machine and can result in an even larger problem than the clustered one.
On the positive side, the re-optimisations for each cluster can be run in parallel.

\subsection{Minimal Excess Electricity} \label{sec:minexcess}

Our third Ansatz for disaggregation is motivated by finding a compromise in terms of computational resources. It is not evident that solving the full optimisation problem is necessary to distribute renewable capacity. Instead, a simpler new objective function that minimises (renewable) excess electricity is defined. It is designed to spatially align renewable generation with demand and possible flexibility options inside the cluster:
\begin{align}
	\label{eq:minexcess_objective}
	&  \underset{G_{v,s}}{\mathrm{min}} \sum_{{v\in \mathcal{V}_c, \atop  s\in S^{\mathrm{re}}},\atop t\in\mathcal{T}} \Bigl[ \bar{g}_{v,s,t}G_{v,s} - d_{v,t} - 0.7 \sum_{{l_{(v,w)}\in\mathcal{L}:\atop  v=c \lor w=c}} F_{(v,w)} \Bigr]^+.
\end{align}
The bracket $\bigl[x\bigr]^+\coloneq \mathrm{max}\{0,x\}$ yields the positive part of the sum.
Compared to the regional re-optimisation introduced in Section \ref{sec:reopt}, the
set of additional constraints to the optimisation problem is much smaller. Only
Eqs.~(\ref{eq:uniform_constraints,eq:reopt_eqconstraint}) are imposed. The
choice of this objective function is motivated by prior research carried out
by~\cite{energyPLAN} or~\cite{Frysztacki2020} where Eq.~(\ref{eq:minexcess_objective}) was invoked as a measure to balance the spatial
resolution of the model and accurate modelling results.
Similarly to the previous methods, this disaggregation method can be run in parallel for each cluster.

\subsection{Modelling Power Flows between Clusters} \label{sec:powerflows}

For two of the three proposed disaggregation methods (`re-optimize' and `min excess'), additional boundary constraints on the inter-cluster transmission lines are added to simulate electricity im- and exports. This can be done by extracting the optimal power flows of the low-resolved network $f_{(c,d),t}$ and distributing them proportional to the capacities of the inter-cluster highly-resolved transmission lines $F_{(v,w)}$ following
\begin{align}
	\label{eq:powerflow_fine}
	& f_{(v,w),t} =\frac{F_{(v,w)}}{F_{(c,d)}}f_{(c,d),t} \in \bigl[-F_{(v,w)}, F_{(v,w)}\bigr] \\
	&  \quad \forall (v,w)\in {E}: v\in \mathcal{V}_c, w\in\mathcal{V}_d\land c\neq d\,. \nonumber
\end{align}
The resulting fine power flow $f_{(v,w),t}$ is modelled as additional demand imposed on all nodes $v\in \mathcal{V}_c$ and $w\in \mathcal{V}_d$ that are connected by a transmission line $(v,w)\in E$, with $c\neq d$, i.e.~\begin{align*}
	d_{v,t} &\mapsto d_{v,t} + f_{(v,w),t} \quad \forall v\in \mathcal{V}_c\,, \\
	d_{w,t} &\mapsto d_{w,t} - f_{(v,w),t} \quad \forall w\in \mathcal{V}_d\quad (c\neq d)\,,
\end{align*}
where positive power flows represent electricity imports and negative ones electricity exports. These results do not deviate strongly from those where each region is treated as an island, meaning that no powerflows retrieved from the coarse model are considered in the disaggregation. However it is plausible to include them when inverting modelling results, therefore this article focuses on this approach. Islanded results can be found in the Appendix, see \ref{appendix:island}.

\subsection{Study Design} \label{sec:studydesign}

To investigate the quality of the proposed disaggregation methods, the model is solved as a pure operational problem, where no further capacity can be built. This is equivalent to solving Eq.~(\ref{eq:objectivefunc}) with its associated constraints, however the technology capacities $G_{v,s}$ are removed from the set of optimisation variables and replaced by a fixed number that is the result of the disaggregation. Load-shedding generators with high but non-extendable capacity are added to the network to guarantee physical feasibility. The operational problem is computationally less extensive to solve because the inter-temporal capacity expansion has been removed from the problem. Therefore, solving a spatially highly-resolved model becomes computationally feasible.

The amount of load-shedding in the highly-resolved operational network model with the disaggregated investment variables taken from the low-resolved capacity expansion model, and the amount of renewable curtailment are considered as main quality measures. Curtailment describes how much abundant electricity the low-resolved model chooses to generate which, when highly-resolved, cannot be transported to locations with high electricity demand. This is mostly due to an inaccurate choice of siting capacity due to missing information about possible transmission bottlenecks in the low-resolved model. Load-shedding is chosen because it indicates how much capacity is underestimated by the low-resolved model due to averaging capacity factors and removing grid bottlenecks from the network. Load-shedding could stem from different reasons: (i) the disaggregation of solar and wind capacities to multiple sites with different capacity factors could result in a lower overall yield compared to the aggregated site, or  (ii) the grid bottlenecks inside the clusters could cause congestion, such that power generated at locations with surplus of electricity cannot be transported to locations with high net load.

As load-shedding is a greater risk in terms of energy security, a test is designed to better understand its origin. To rule out reason  (ii), a second operational scenario is run (which is referred to as ``copper-plate''), where the capacity of all  transmission lines that have vanished in the low-resolved network due to aggregation are set to $\infty$, i.e.~\begin{equation}
	F_{(v,w)}\rightarrow\infty \quad \forall (v,w): v,w \in \mathcal{V}_c, \ \forall c\,. \label{eq:infinitetransmission}
\end{equation}
This modification is only applied to solve the operational problem, not for the disaggregation of results.

Note that all inter-cluster transmission capacity is still finite, meaning
\begin{equation}
	F_{(v,w)} \ll \infty \quad \forall (v,w): v \in \mathcal{V}_c,\, w\in \mathcal{V}_d,\, c\neq d\,. \label{eq:finitetransmission}
\end{equation}

A summary of the two considered scenarios (``regular'' and ``copperplate'') is provided in Table
\ref{tab:intra-scenarios}.

\begin{table}
	\caption{Investigated intra-cluster scenarios for each of the disaggregation methods. This means that additional constraints on the transmission lines within each cluster are formulated.}\label{tab:intra-scenarios}
	\begin{tabular}{@{} lp{5.8cm}}
		\toprule
		\textbf{Short name} & \textbf{Scenario description} \\ 
		\midrule
		\ & \hspace*{-6.3em} The highly-resolved network with disaggregated capaci- \\
		\ & \hspace*{-6.3em} ties is solved as an operational problem where ...  \vspace*{-.5em}\\
		\textbf{regular} & ... no further adaptations are made. \\
		\textbf{copperplate} & ... the intra-cluster transmission capacity is infinitely high. Note, that the inter-cluster transmission capacity is still bound. \\
		\bottomrule
	\end{tabular}
\end{table}

\section{Results}\label{sec:results}

We first present the feasibility of low-resolution modeling results when dis-aggegated into high spatial resolution in section \ref{sec:methodfeasibility} using the three proposed dis-aggregation methods. We distinguish between the ``regular'' set-up where the intra-cluster transmission capacity is not changed (section \ref{sec:resultsregular}) where we additionally analyze where the curtailment and load-shedding measures are spatially located (section \ref{sec:localisation}) and the ``copperplate'' one, where the transmission capacity within clusters of the spatially highly-resolved model is set to infinity to approximate the clustered copperplate optimisation model (section \ref{sec:resultscopperplate}). Then, we discuss computational Trade-Offs of the presented dis-aggregation methods in section \ref{sec:tradeoffs}.

\subsection{Feasibility Considerations} \label{sec:methodfeasibility}

In this section the feasibility of the spatially highly-resolved operational model is discussed that consists of the disaggregated spatially low-resolved modelling investment variables. Here, ``feasibility'' means the notion that electricity demand can be covered by 100\% renewable electricity and no additional generation of conventional plants is necessary for a stable operation of the grid. Finally, the model is modified to study possible reasons of necessary intervention measures to secure electricity supply at all times and places.

\subsubsection{Regular Intra-Cluster Transmission Capacity} \label{sec:resultsregular}

\begin{figure}
	\centering
	\includegraphics[width=.235\textwidth]{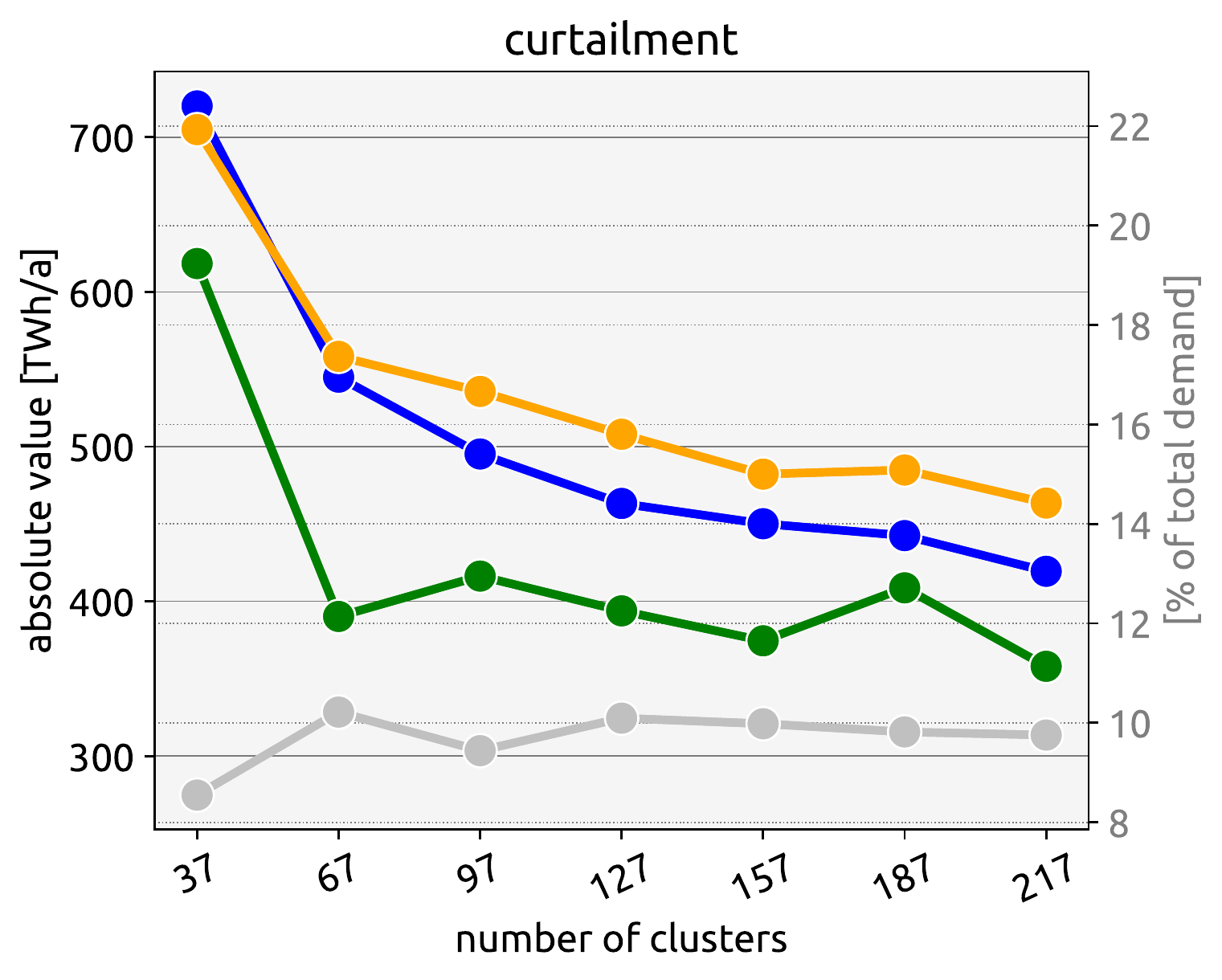}
	\includegraphics[width=.235\textwidth]{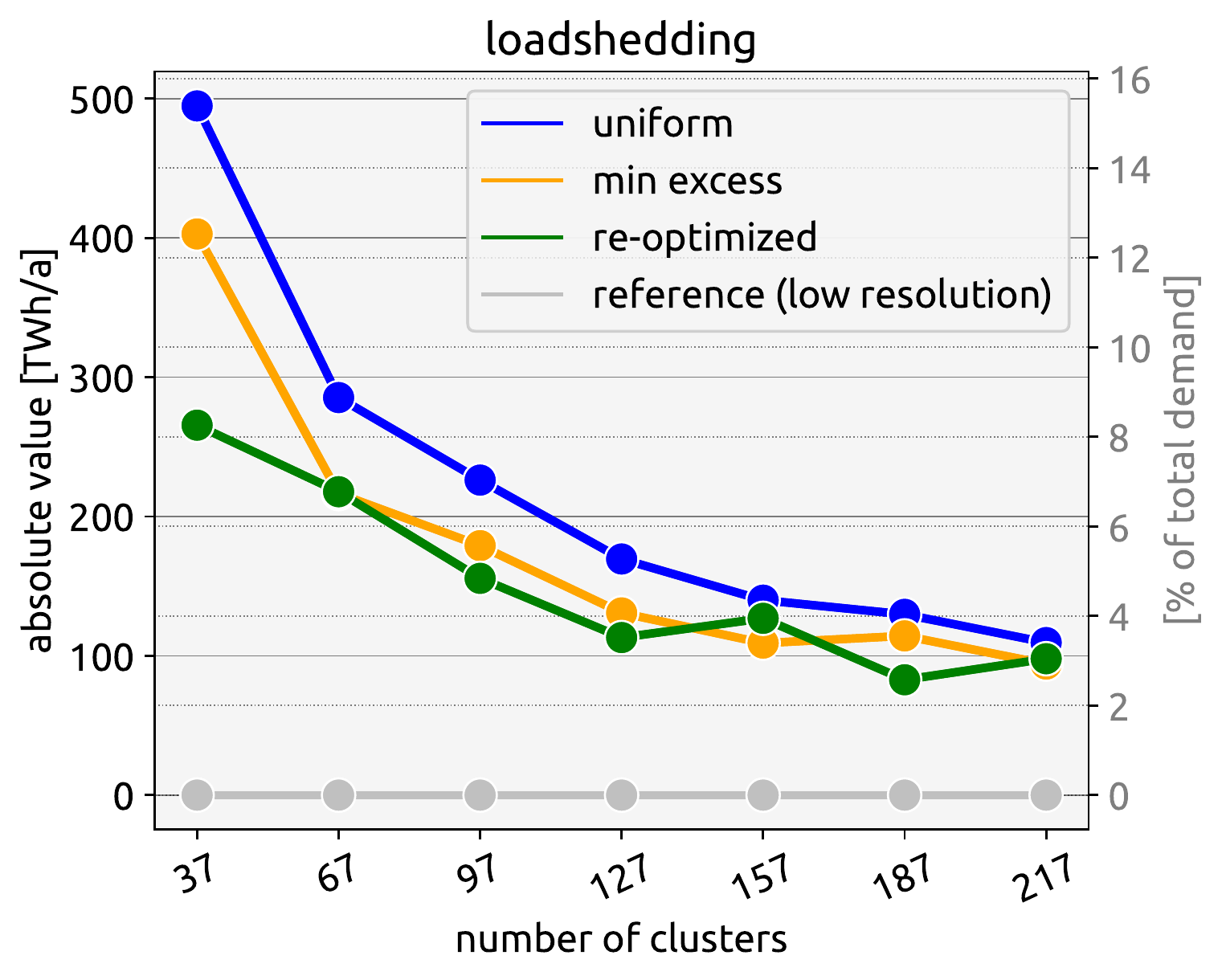}
	\caption{Amounts of annual load-shedding and curtailment after solving the high-resolution operational problem with the three proposed dis-aggregation methods. Transmission capacity within every cluster is not adjusted. (``regular'' setup in Table \ref{tab:intra-scenarios}).}\label{fig:loadsheddingandcurtailment_plain_imports}
\end{figure}

The results presented here are derived from the ``regular'' set-up, meaning with no adjustment to the intra-cluster grid capacities in the high resolved optimisation model (see Table
\ref{tab:intra-scenarios}). The resulting amounts of load-shedding and curtailment are presented in  Figure \ref{fig:loadsheddingandcurtailment_plain_imports}.

For any of the three proposed disaggregation methods it can be seen that the amounts of both curtailment and load-shedding decrease as the resolution of the underlying low-resolved capacity expansion model increases. This can be explained by a better approximation of the spatially highly-resolved model. The higher the spatial resolution of the capacity expansion model, the better is its approximation of the original highly-resolved model. Therefore, at higher spatial resolution, less costly load-shedding measures are necessary when disaggregating investment variables. It can also be seen that load-shedding is caused by high curtailment rates, that are likely to be provoked by transmission congestion. However, there are substantial differences in the performance of the disaggregation methods.

Curtailment rates of the different disaggregation methods deviate by $1\text{--}3\%$ from one another on average, depending on the underlying capacity expansion model resolution. Distributing coarse investment variables across the spatially highly-resolved operational model using the `min excess' approach yields the highest curtailment rates of $11\%\text{--}22\%$ of annual electricity demand, depending on the low-resolved capacity expansion model resolution and the disaggregation method. The lowest resolution has the highest curtailment. Uniformly distributing results performs similar to the `min excess' method at a very low resolution of $37$ nodes (one node per country), resulting in 22\% of curtailed electricity. The `re-optimise' method performs better in this regard, resulting only in 19\% of curtailed electricity. But the curtailment rates decrease to approximately 14.5\% (`min excess'), 13\% (`uniform') and 11\% (`re-optimised') of annual electricity demand, as the capacity expansion model resolution increases. Re-optimising the local problem yields the lowest curtailment for every low-resolved model resolution, which is approximately $2\text{--}3\%$ lower compared to the results of the `uniform' approach.

Regarding load-shedding, for a low-resolved capacity expansion network where every country is represented by a single node ($37$ clusters in Figure \ref{fig:loadsheddingandcurtailment_plain_imports}), `re-optimise' performs best as it results in the lowest load-shedding rates. Re-optimising yields approximately $265$~TWh or 8.2\% of the annual electricity demand that cannot be covered by renewable generation. If this gap were filled with gas to satisfy electricity demand, annual carbon emissions would rise from 0\% to 3.2\% of 1990s levels.
Compensating the unmet demand when disaggregating results with the `min excess' method yields approximately $400$~TWh of load-shedding, resulting in 4.8\% of carbon emissions (1.6\% more compared to `re-optimised') if gas is dispatched for the load-shedding measure.
When uniformly disaggregating renewable capacity within the clusters, the operational problem returns $500$~TWh of load-shedding measures. Compensating with gas would result in 6\% of carbon emissions of 1990, 1.2\% more compared to `min excess'.

When increasing the capacity expansion model resolution, the amount of load-shedding decreases for all the disaggregation methods. It can be seen that at a model resolution of $67$ or more nodes, the amount of load-shedding is in the same range for the methods `re-optimised' and `min excess' deviating by only 0.5\% on average. When uniformly distributing the retrieved low-resolved, optimal capacities, load-shedding measures are higher than those of the competing disaggregation methods by initially 2.8\% at a capacity expansion resolution of $37$ nodes and linearly decreases as the resolution of the capacity expansion model increases. At around $187$ nodes, the difference for all three methods is below 0.5\% in terms of necessary load-shedding measures. At an underlying capacity expansion model resolution of $217$ nodes, the amounts of load-shedding are all within the range $94\text{--}110$~TWh, corresponding to $3\text{--}3.5\%$ of annual electricity demand in Europe.

\subsubsection{Localisation of Load-Shedding and Curtailment} \label{sec:localisation}

\begin{figure} \centering
	\includegraphics[width=.48\linewidth]{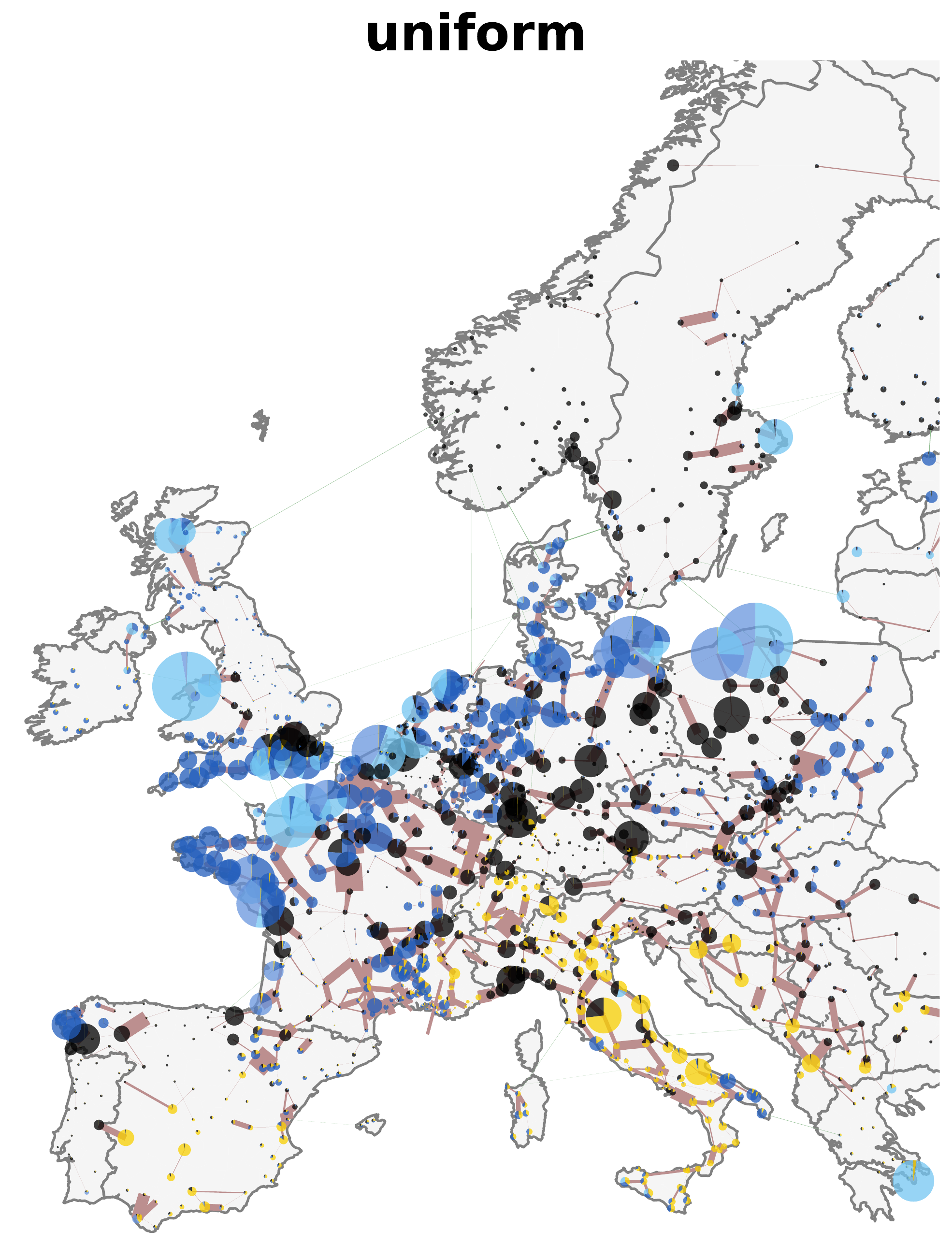}
	\includegraphics[width=.48\linewidth]{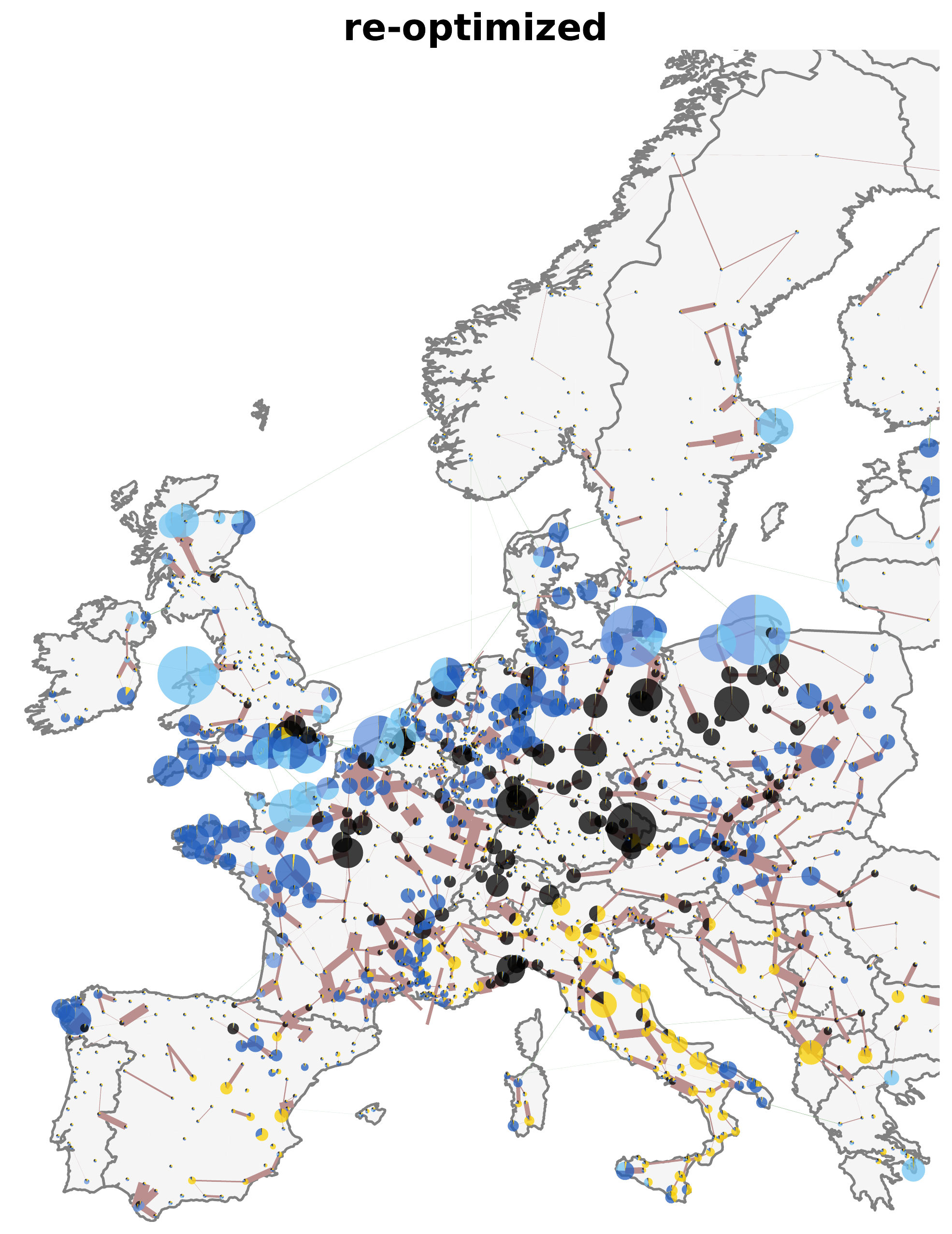}
	\includegraphics[width=.48\linewidth]{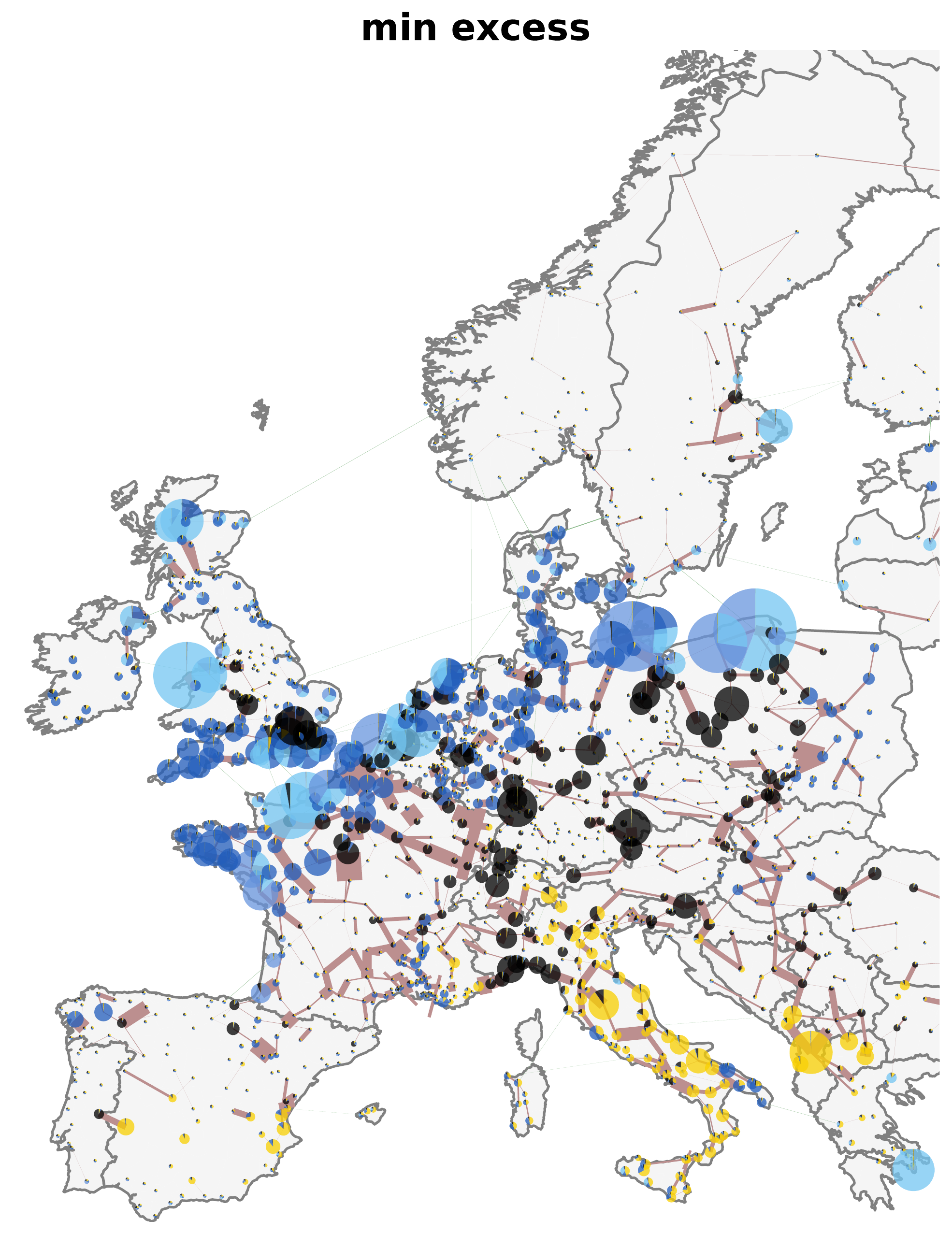}
	\includegraphics[trim=2em 2em 20em 2em, clip, width=.48\linewidth]{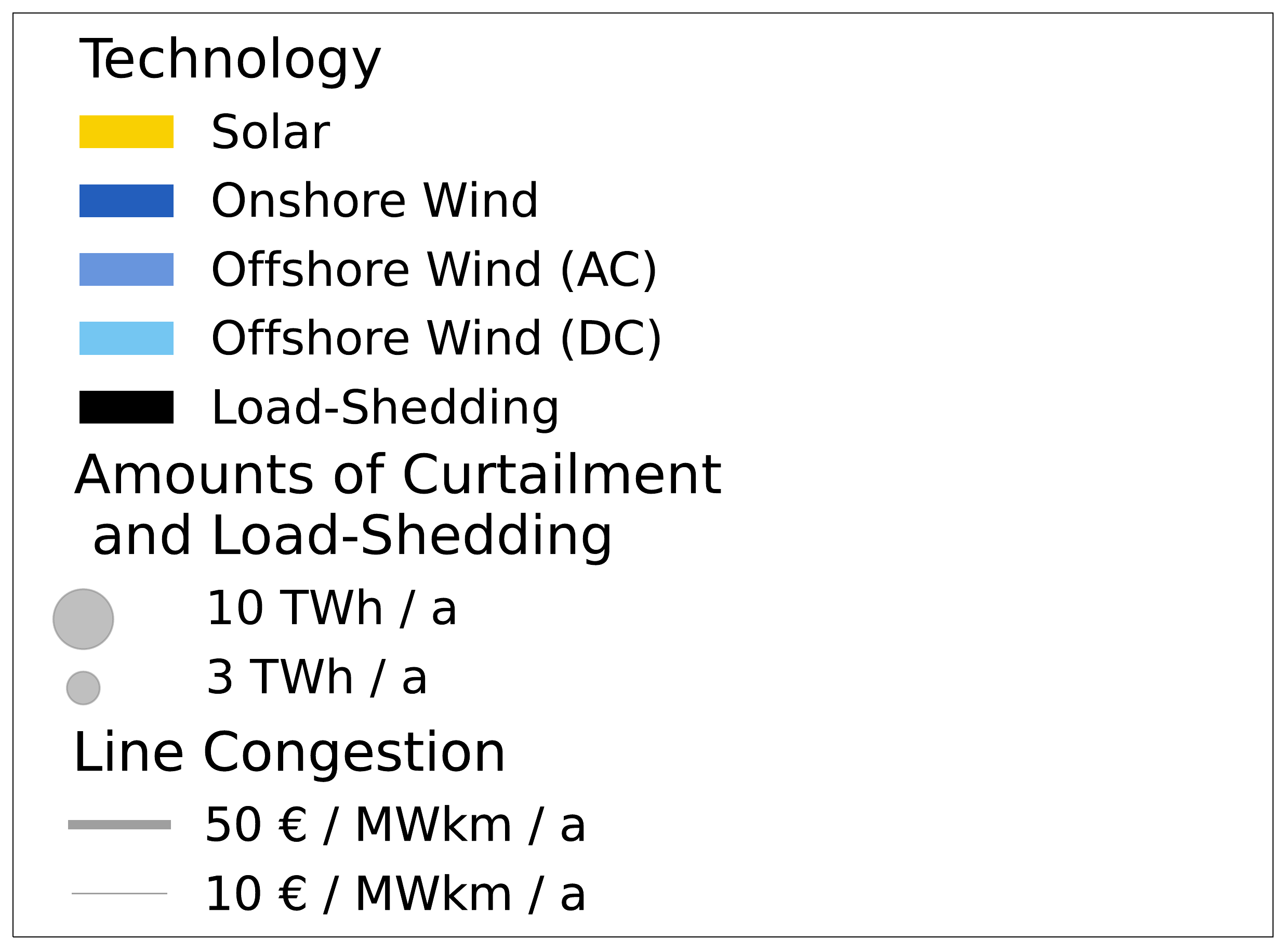}
	\caption{Spatial distribution of curtailment and load-shedding measures across Europe in the highly-resolved operational $1250$ node model for the three proposed disaggregation methods after running the operational dispatch. Optimal capacity installations are taken from a model resolution of $97$ nodes, and are disaggregated at $1250$ regions.\label{fig:spatial_load}}
\end{figure}

In this section it is analysed where curtailment and load-shedding is spatially localised. Recall that the highly-resolved model yields load-shedding measures because of  (i) disaggregating capacity factors results in a different overall yield of renewable electricity or  (ii) grid bottlenecks that did not occur in the low-resolved model, as described in detail in Section \ref{sec:studydesign}.

Figure \ref{fig:spatial_load} displays the regions of curtailment and load-shedding spatially distributed after running the operational highly-resolved $1250$ node model for all three disaggregation methods for a reference capacity expansion model resolution of $97$ nodes. In all three cases it can be seen that the load-shedding is scattered in central European regions such as southern Poland, central and southern Germany, Switzerland and Austria and thus far from coastal areas and southern locations, such as e.g.~northern Germany and France, Italy and Spain. At the same time, coastal and southern locations have high amounts of curtailment. Transmission lines connecting regions with high amounts of curtailment and regions with high load-shedding show high congestion rates. Thus, the results suggest that the low-resolved capacity expansion model favours investments in wind turbines at locations with good wind conditions at coastal areas, and in solar panels in the southern regions with good solar radiation, while it is blind to transmission bottlenecks that prohibit transporting the electricity to demand centres.

\subsubsection{Infinite Intra-Cluster Transmission Capacity} \label{sec:resultscopperplate}

\begin{figure}
	\centering
	\includegraphics[width=.235\textwidth]{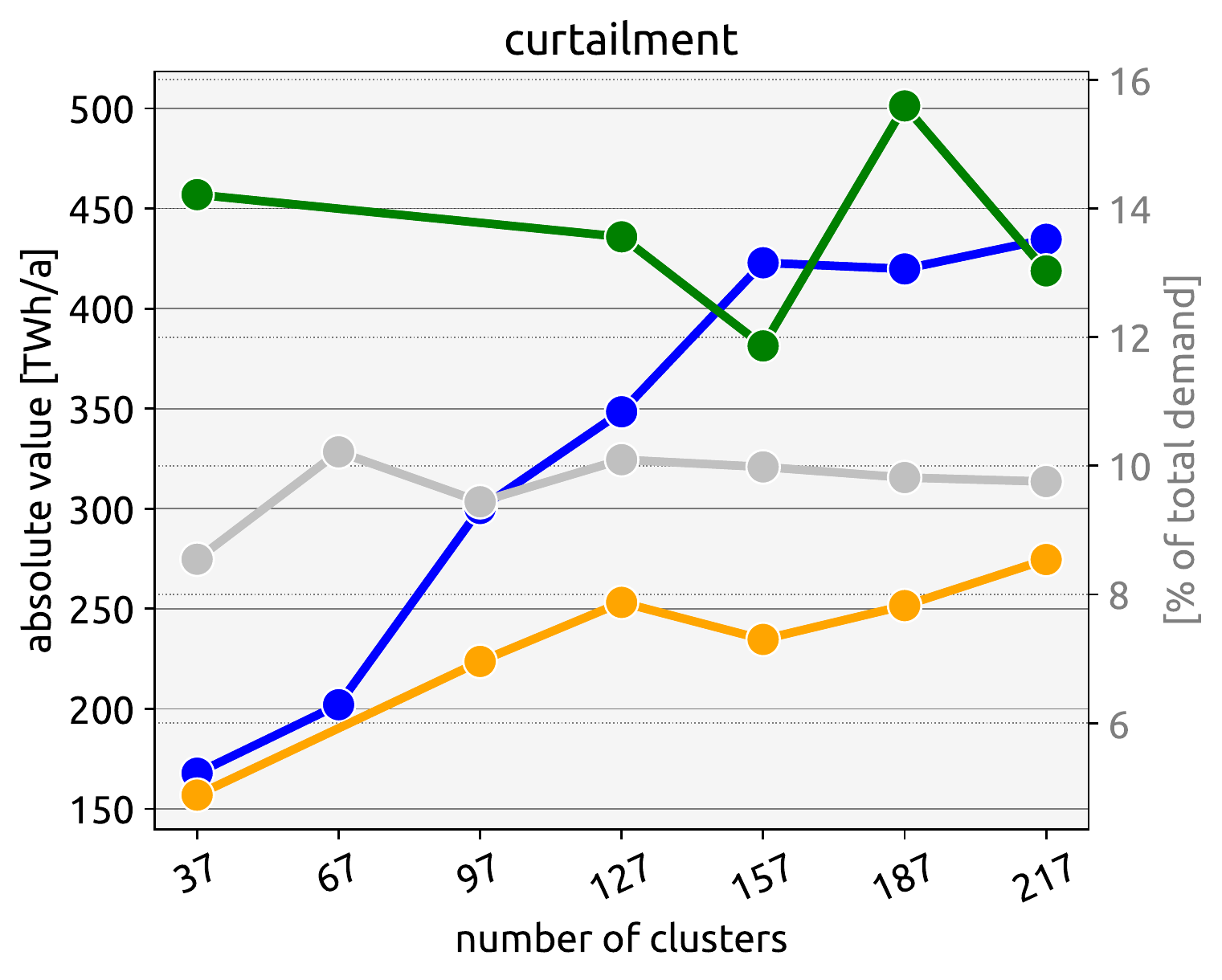}
	\includegraphics[width=.235\textwidth]{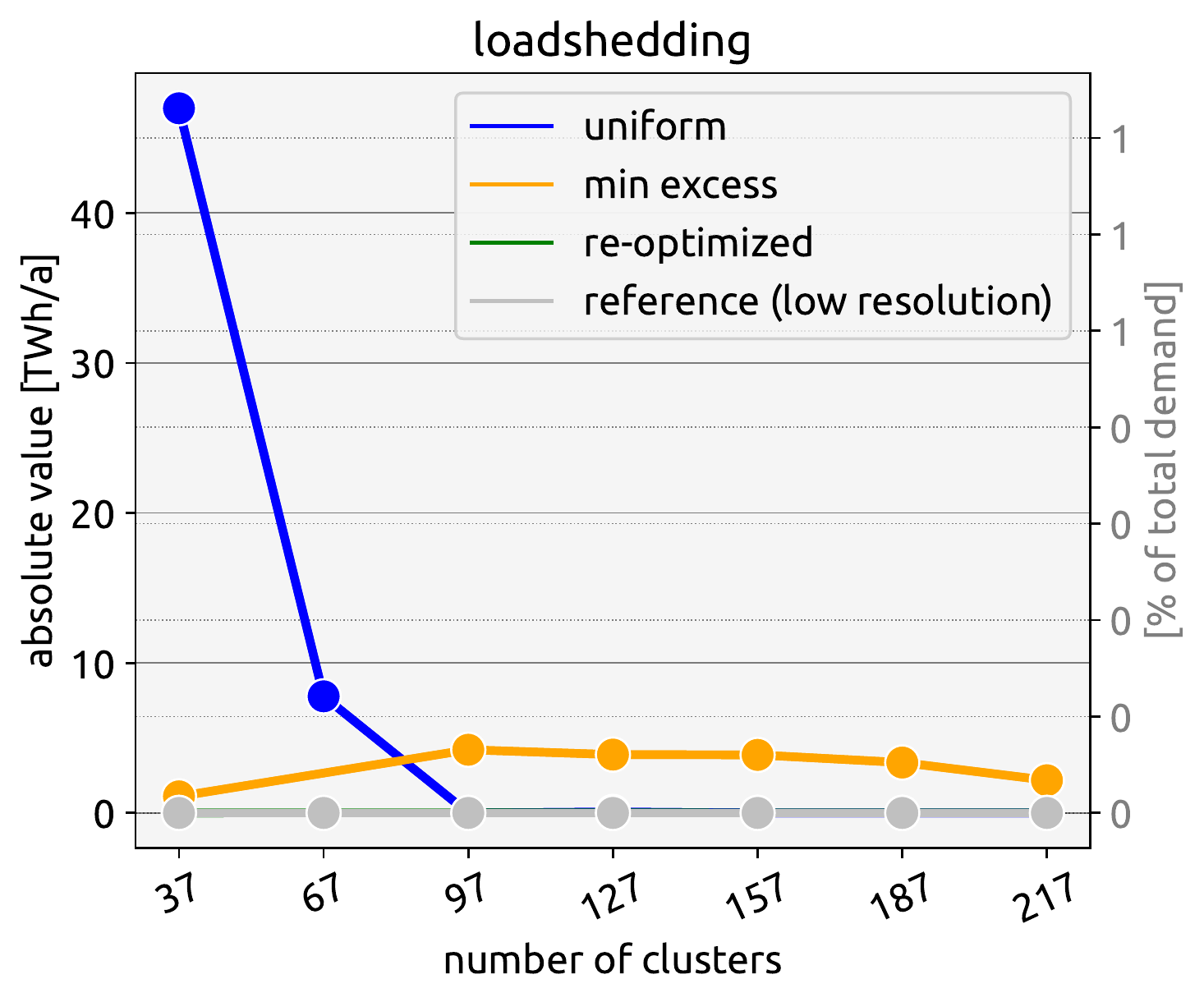}
	\caption{Amounts of annual load-shedding and curtailment after solving the highly-resolved operational problem with the three proposed disaggregation methods. The number of clusters on the $x$-axis refers to the spatial resolution of the model of which the investment variables are disaggregated at high resolution. Transmission capacity within every cluster is set to infinite to approximate the copperplate clustered network (``copperplate'' set-up in Table \ref{tab:intra-scenarios}).\label{fig:loadsheddingandcurtailment_copperplate_imports}}
\end{figure}

To verify why load-shedding measures are necessary as well as to better understand the high curtailment rates, a setting is considered where within each cluster the transmission capacity is set to infinity, following the description provided in the beginning of Section \ref{sec:studydesign}, see Eqs.~(\ref{eq:infinitetransmission})--(\ref{eq:finitetransmission}). This means that in the highly-resolved model, only the capacity between clusters is limited. Results on load-shedding and curtailment are presented in Figure \ref{fig:loadsheddingandcurtailment_copperplate_imports}.

It can be seen that the amounts of renewable curtailment in the disaggregated operational models deviate by less than 5\% from the curtailment rates of the capacity expansion model. They can mainly be explained by varying capacity factors. In the highly-resolved operational models, larger deviations of capacity factors within each clustered region become available compared to the spatially low-resolved capacity expansion model.

As the resolution of the capacity expansion model increases, the amount of curtailment also tends to increase slightly. This can be explained by the fact that more total capacity is installed for a high  capacity expansion model resolution.

In terms of congestion, the necessary amount of load-shedding when uniformly disaggregating renewable capacity drops to $0$ for a capacity-expansion model resolution of above $100$ nodes. For a one-node-per-country model ($37$ nodes), there remains a relatively low amount of load-shedding of approximately $50$~TWh, resembling about 1.5\% of annual electricity demand.
In case the `re-optimise' disaggregation Ansatz is invoked, load-shedding decreases to 0\% of annual electricity demand for every low-resolved capacity expansion model.
`min excess' yields less than $5$~TWh (${<}0.5\%$) of load-shedding measures for any low-resolved capacity expansion model. At peak ($97$ nodes)
this amount of gas would
emit around $800$ kg of CO$_2$ (0.05\% of 1990s emissions).
One can conclude that these results are consistent with the main cause of load-shedding being the transmission restrictions within the clusters.

\subsection{Trade-Offs of the Dis-Aggregation Approaches} \label{sec:tradeoffs}
There are four main qualities that can be considered when evaluating trade-offs of the different disaggregation methods. First, the quality of results: How well do the proposed methods solve the problem at hand? Second and third, the computational efforts can be considered. These mainly focus on the question: Are the proposed methods computationally legitimate for the considered problem? This consideration includes not only the memory requirements needed to solve the problem, but also the time it takes to solve. Fourth, depending on the results of the methods or the problem formulation, it might also be worth considering the efforts to implement a solver.

For the proposed methods in this paper, a summary of these four qualities is provided in 
Table \ref{tab:tradeoffs}.

\begin{table}
	\caption{Trade-Offs of the three proposed approaches to dis-aggregate results. Marked with a \cmark\ indicate a reasonable trade-off, entries with a \xmark\ indicate an inadequate compromise.}\label{tab:tradeoffs}
	\begin{tabular}{l | c c c c}
		\textbf{} &
		\rotatebox[origin=c]{90}{\textbf{Implementation}} &
		\rotatebox[origin=c]{90}{\textbf{Solving Time}} &
		\rotatebox[origin=c]{90}{\textbf{Memory (RAM)}} &
		\rotatebox[origin=c]{90}{\textbf{Results Quality}} \\
		\midrule
		\textbf{uniform} & \cmark & \cmark & \cmark & \xmark \\
		\textbf{min excess} &
		\xmark & \cmark & \cmark & \cmark \\
		\textbf{re-optimize} &
		\xmark & \xmark & \xmark & \cmark \\
		\bottomrule
	\end{tabular}
\end{table}

The performance with respect to the quality of results of the proposed disaggregation methods was already discussed in Section \ref{sec:methodfeasibility}. Now, the performance of the proposed methods is analysed from a computational point of view. Rating the efforts of implementation is a subjective task, therefore it is solely related to the fact that uniformly distributing a number across a set of nodes does not involve mathematical optimisation. Therefore applying an uniform distribution is rated ``easier'' than formulating a mathematical constraint to an existing optimisation problem as proposed in `re-optimise', or a whole optimisation problem including both objective function and associated constraints, as proposed in `min-excess'.

\begin{figure}
	\centering
	\includegraphics[width=.45\textwidth]{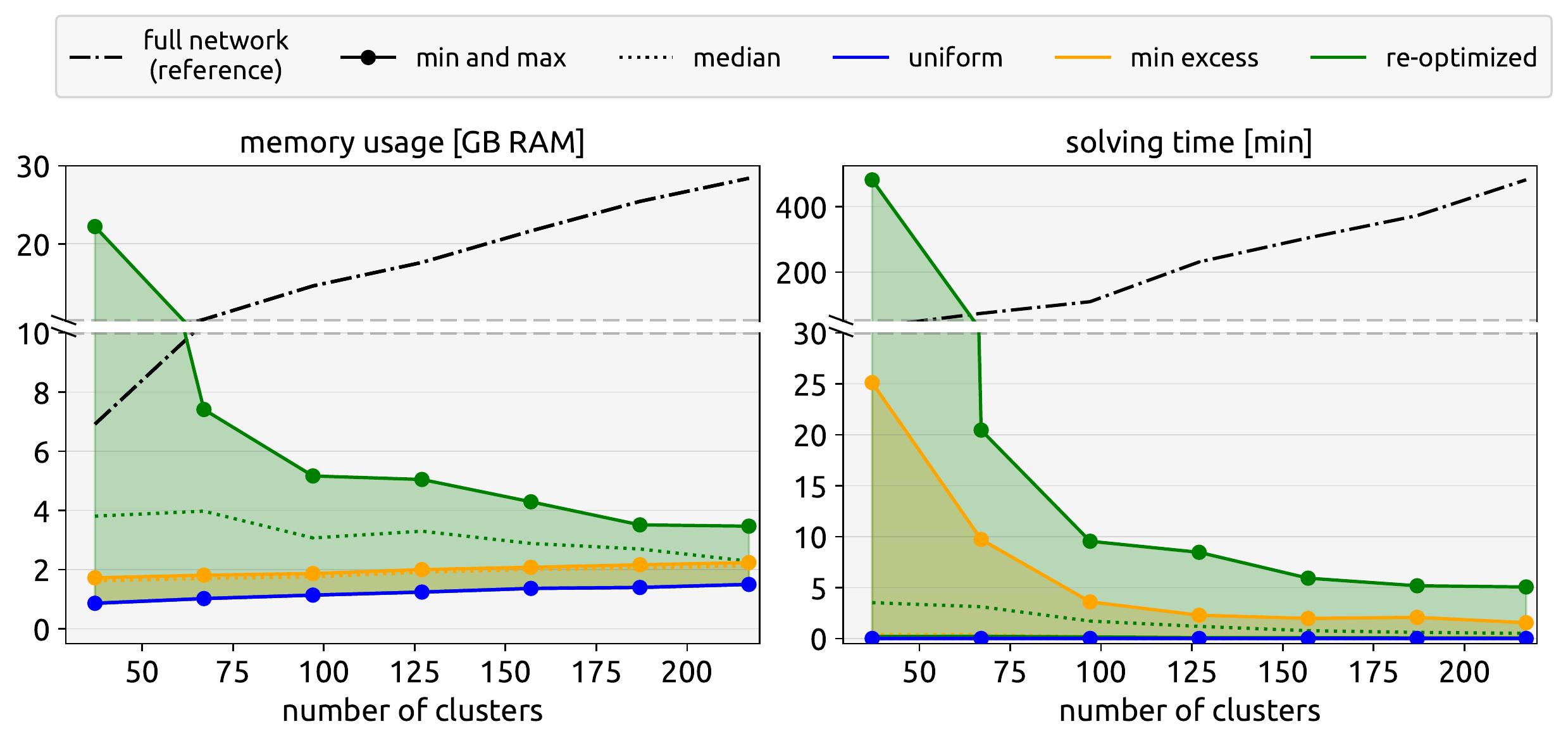}
	\caption{Memory resource requirements (left) and solving times (right) for executing the proposed disaggregation methods for individual regions. The number of clusters on the $x$-axis refers to the spatial resolution of the model of which the investment variables are disaggregated at high resolution. Possible memory consumption and solving times are marked by the corresponding colour shades, depending on the size of the cluster that is to be disaggregated. The memory resource requirement and solving time for the largest and smallest local problem can be taken from the edges of the marked area. Median requirements are displayed with a dotted line. The black dotted line denotes the requirements of solving the spatially low-resolved network that has the spatial resolution as given on the $x$-axis. Note that for `min excess' and `uniform', the median values coincide with the lower bounds, and are thus hidden in the plot.  Moreover, for `uniform', there upper and lower bounds collapse, thus no distribution is visible.\label{fig:tradeoffs}}
\end{figure}

Computational resources and solving times for disaggregating spatially low-resolved model results at high spatial resolution are presented in Figure \ref{fig:tradeoffs} for every proposed method.

Resource-wise, re-optimising the local model consumes up to 13 times (1.7 times in average) the amount of resources compared to minimising a simpler objective in ``min excess'', and up to 26 times (2.7 times in average) compared to uniformly distributing the capacity obtained from the low-resolved model (``uniform''). In absolute numbers, the method ``re-optimise'' consumes up to 22.2~GB RAM at peak, compared to 2.2~GB RAM for ``min excess'' and only 1.5~GB RAM in case of ``uniform''. Today`s average state-of-the-art personal computers are able to solve both the ``uniform'' and ``min excess'' problem formulations for any model resolutions, while solving the ``re-optimise'' approach needs more computational power and, therefore, requires a more advanced machine or even a high-computational cluster access. All local disaggregation runs were carried out in parallel.

Considering the computational times, these trade-offs are similar. The method ``uniform'' is up to $4000$ times faster at peak than ``min excess'' and $70$ times faster in average. In turn, ``min excess'' is up to $20$ times faster than ``re-optimise'' and $11$ times faster in average. Note that computational times might change when allowing a lower accuracy of the results. Here, a barrier convergence tolerance of 10$^{-9}$ is chosen and a feasibility tolerance of 10$^{-6}$, which is not necessarily required. A tolerance of 10$^{-3}$ might suffice in most applications. However, lowering the tolerance of the solver reduces solving times, but the memory consumption persists.

All experiments presented in this article were carried out on a high-computational cluster with 5 nodes, each having an allocatable capacity of 48 cpu's and 256~GB memory.

\section{Conclusions}\label{sec:conclusions}
From these results, several conclusions on the methodology of the disaggregation methods can be drawn as well as on the insights of disaggregating coarse modelling results at a higher spatial detail.

The presented methods to disaggregate optimal infrastructure investment of renewable generation technologies and flexibility options have significant differences in their quality of results, simplicity of implementation and computational resource consumption. It has been shown that it is not necessary to locally solve the full optimisation problem to disaggregate coarse results at higher spatial detail, as it was conducted in previous research. Instead, it can be sufficient to formulate a suitable alternative objective which reduces computational cost and is able to preserve the quality of the disaggregation. In this paper, a novel function ``min excess'' has been suggested that performs just as well, for lower computational burden. Further inverse functions could be considered in future research.

Regarding the disaggregated highly-resolved modelling results, the result presented in this article once again stress that modelling a fully renewable European electricity system at a resolution of one node per country is insufficient to retrieve reliable capacity expansion suggestions.
Moreover, results retrieved from models that simulate a fully renewable electricity system that are clustered to a spatial resolution of around $100\text{--}200$ nodes using state-of-the-art evaluated aggregation methods fail to cover approximately $100$~TWh of Europe's electricity demand, approximating $3\text{--}5\%$ of its annual consumption. Instead of consuming the excess electricity, curtailment rates rise by approximately the shed amount, additional to what would have been expected for an economic optimum.
Our analysis reveals that the electricity shortage is due to local transmission constraints. Spatially low-resolved models assume that power can be transferred without limit to all locations that are represented within a single region. Therefore, intra-nodal transmission constraints are ignored in the aggregated model. Thus, disaggregated results at higher spatial detail are confronted with power flow restrictions, resulting in transmission congestion and imply necessary load-shedding measures, eventually making the investment decisions retrieved from a coarse model sub-optimal and technically infeasible, if no additional investments can be assumed. These findings imply that accurately representing transmission and power-flows in the model is of high relevance to find a cost-optimal or low-cost solution that is technically feasible. Our results do not show that a fully renewable system is not possible. Conversely, the insight from our results together with lessons learned from spatial clustering studies provide valuable insights that emphasise the relevance of high resolution modelling, such that a fully renewable system can be achieved at low cost.

\section{Limitations of this Study} \label{sec:limitations}
Removing the set of optimisation variables that accounts for the capacity expansion allows solving an operational dispatch model at a higher model resolution. Nevertheless, due to a persisting computational burden, the presented operational model results are based on model runs retrieved from a model resolution of 1250 nodes, i.e.~approximately 25\% of the original network size. Therefore, the resulting amounts of load-shedding and curtailment resulting in the disaggregated operational model runs are likely to increase if the operational model was spatially higher resolved (for example at 5000 nodes), and, thus, strengthen our main argument.

This study analysed methods to disaggregate spatially low-resolved optimal generation variables. However, the study did not investigate methods to disaggregate transmission capacity expansion modelling results, or how additional transfer capacity obtained from a transmission expansion problem could improve the overall results. Such an analysis could build on our presented methods and extend them on an additional optimisation variable.
Moreover, all results presented in this paper were carried out for a fully self-sufficient and fully renewable Europe. Lowering the carbon emission target could relax the findings and would not make as strong implications. Therefore, in a future study, different carbon emission targets could be analysed more carefully.

Results of this study are all based on the MIT-licensed models PyPSA v0.18.0 and PyPSA-EUR v0.3.0. Therefore, nearly all of the limitations that apply for this version of the model also apply for this study. These include for example retrieving optimal capacities that rely on weather data from a single weather year, applying only a linearised power flow model or neglecting dynamic line rating. Some of these simplifications might improve in future model releases.

\section{Data Availability}\label{sec:data}
All code is or will be publicly available on github under an open source license.

\section{Glossary}\label{sec:glossary}
Abbreviations and variables are documented in Table \ref{tab:glossary}.
\begin{table}
	\caption{Glossary. Variables and their description.}\label{tab:glossary}
	\begin{tabular}{@{} lp{5.8cm}}
		\toprule
		\textbf{Abbrev.} & \textbf{Description} \\ 
		\midrule
		$\mathcal{V}$ & Set of all nodes contained in the model.\\
		$E$ & Set of all edges representing transmission lines contained in the model.\\
		$v, w$ & Representative names for high-resolution nodes.\\
		$c, d$ & Representative names for clustered nodes. \\
		$\mathcal{V}_c$ & Set of highly-resolved nodes that are aggregated to node $c$. \\
		$(v,w)$ & a HVAC or HVDC line connecting nodes $v$ and $w$.\\
		$\mathcal{S}$ & Set of available technologies in the model, for example wind generator or battery storage. \\
		$s$ & Generator or storage technology.\\
		$\mathcal{T}$ & Set of snapshots in the model.\\
		$t$ & snapshot, typically covering a duration of $2$ hours.\\
		$G_{v,s}$ & Capacity in node $v$ of generators of type $s$. \\
		$g_{v,s,t}$ & Dispatch in node $v$ of technology type $s$ at time $t$.\\
		$\bar{g}_{v,s,t}$ & Capacity factor at node $v$ for technology $s$ at time $t$. \\
		$w_t$ & Weighting for time, here $w_t\equiv 2\ \forall t\in\mathcal{T}$, representing a 2-hourly model run.\\
		$\mathcal{L}$ & Set of transmission lines in the model. \\
		$F_{(v,w)}$ & Capacity of the transmission line connecting nodes $v$ and $w$. \\
		$f_{(v,w),t}$ & Power flow from node $v$ to node $w$ at time $t$.\\
		$d_{v,t}$ & Electricity demand in node $v$ at time $t$. \\
		$c_{v,s}$ & Capital costs at node $v$ of technology $s$.\\
		$o_{v,s,t}$ & Operational costs at node $v$ for technology $s$ at time $t$.\\
		$\Upsilon_{V}(t)$, $\Upsilon_{C}(t)$ & Curtailment of the high-resolution model ($V$) or the clustered model ($C$) at time $t$. \\
		$\Delta_{V}(t)$, $\Delta_{C}(t)$ & Load-shedding measure of the high-resolution model ($V$) or the clustered model ($C$) at time $t$. \\
		\bottomrule
	\end{tabular}
\end{table}

\section{Acknowledgements}\label{}

MF and VH acknowedge funding from the Helmholtz Association under the program “Energy System Design”, MF and TB acknowedge funding from the Helmholtz Association under Grant No. VH-NG-1352.

\section{Declaration of Interests}\label{}

The authors declare that they have no competing financial interests.









\appendix
\section{Appendix}\label{}

\subsection{Cost Assumptions} \label{appendix:costs}
Cost assumptions used for the clustered reference models to make cost-optimal investment decisions can be taken from Table \ref{tab:costs}.
\begin{table}\centering
	\caption{Technology investment costs with $1\$ = 0.7532$€.}         \label{tab:costs}
	\begin{tabular}{@{} lrl @{}}
		\toprule
		asset & cost & unit \\
		\midrule
		onshore wind  & 1110 & €/kW \\
		offshore wind & 1640 & €/kW \\
		(AC/DC grid connection separate) & & \\
		solar PV utility & 425 & €/kW \\
		solar PV rooftop & 725 & €/kW \\
		open cycle gas turbine & 400 & €/kW \\
		run of river & 3000 & €/kW\\
		\midrule
		pumped hydro storage & 2000 & €/kW \\
		hydro storage & 2000 & €/kW\\
		battery storage & 192 & \$/kWh \\
		battery power conversion & 411 & \$/kW$_{\textrm{el}}$ \\
		hydrogen storage & 11.3 & \$/kWh \\
		hydrogen power conversion & 689 & €/kW$_{\textrm{el}}$ \\
		\midrule
		HVAC overhead transmission & 400 & €/(MWkm) \\
		HVAC underground transmission & 1342 & €/(MWkm)\\
		HVAC subsea transmission & 2685 & €/(MWkm) \\
		HVDC underground transmission & 1000 & €/(MWkm) \\
		HVDC subsea transmission & 2000 & €/(MWkm) \\
		\bottomrule
	\end{tabular}
\end{table}

\subsection{Results of Island-ed Dis-Aggregation Method} \label{appendix:island}

\begin{figure}
	\centering
	\includegraphics[width=.235\textwidth]{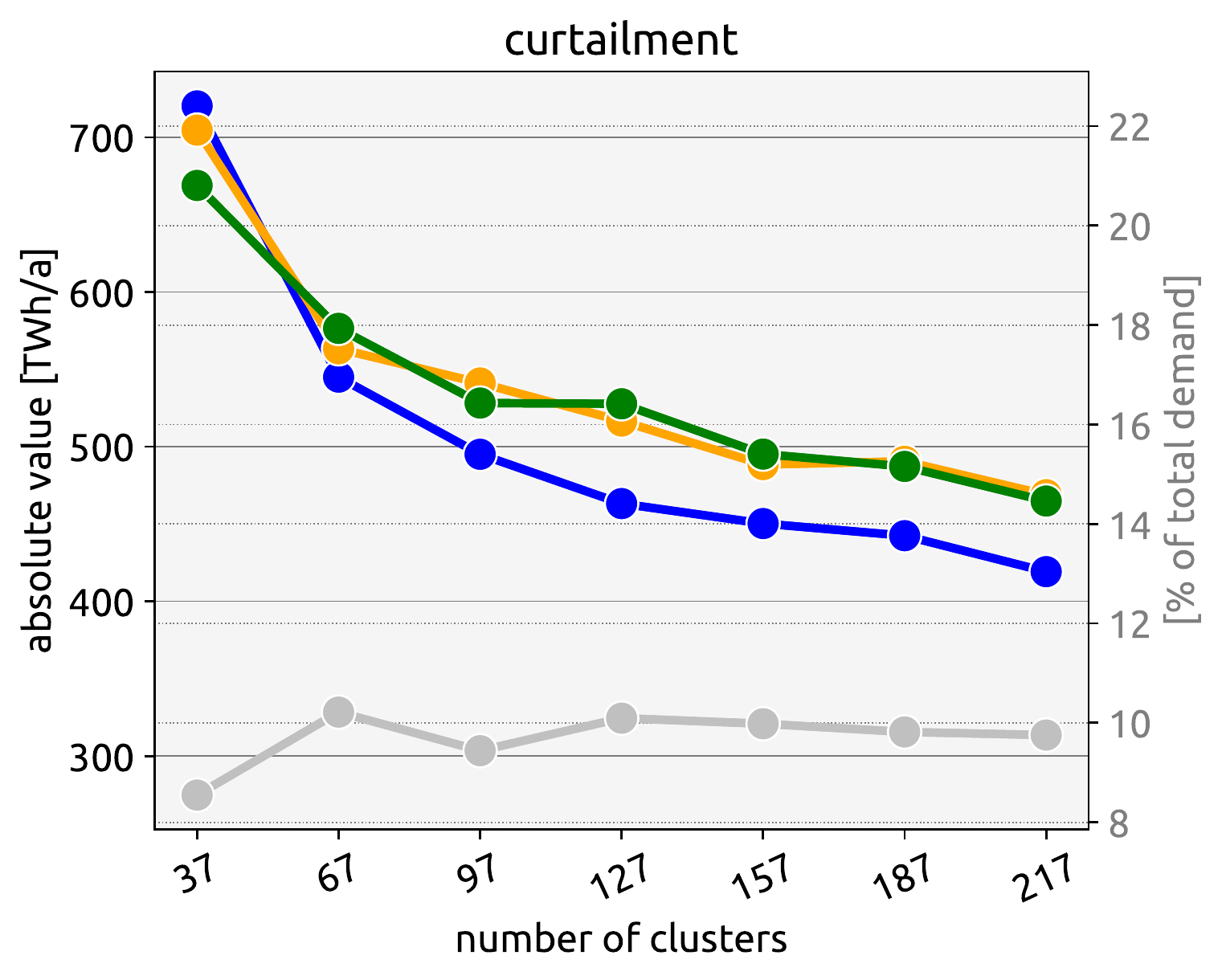}
	\includegraphics[width=.235\textwidth]{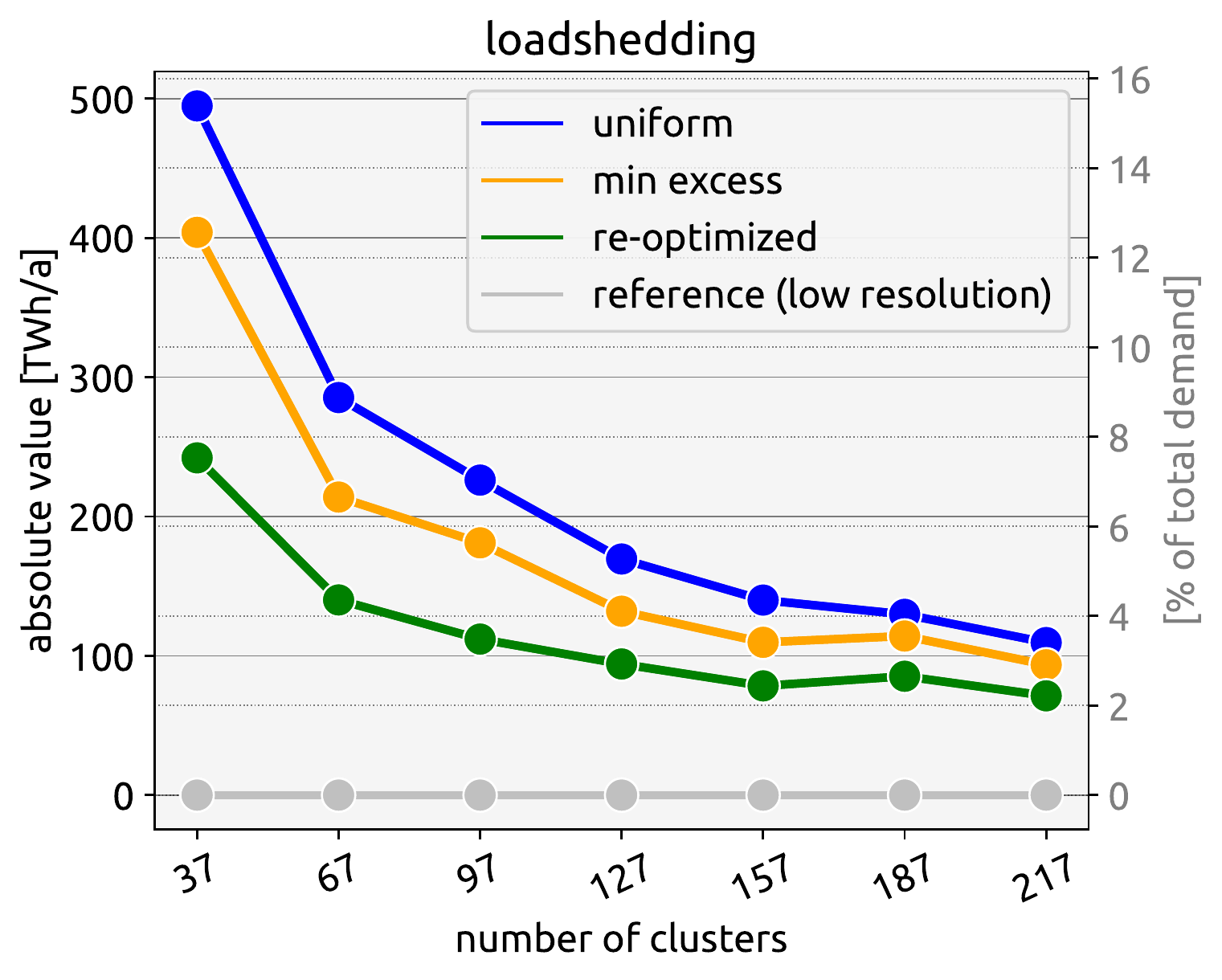}
	\caption{Results as displayed in Figure \ref{fig:loadsheddingandcurtailment_plain_imports} of an island model, meaning that no inter-cluster electricity imports or exports are considered for the disaggregation.\label{fig:loadsheddingandcurtailment_plain_island}}
\end{figure}

In this setting each cluster is treated as an island, meaning that no electricity trade between other clusters is considered for the dis-aggregation. Results on load-shedding and curtailment for this scenario are displayed in Figure \ref{fig:loadsheddingandcurtailment_plain_island}.

The overall trend of the results is similar to the simulations where inter-cluster power flows were considered in the simulations. However, there are minor differences mainly affecting the ``re-optimize'' results. These result in an overall higher curtailment of $2-3\%$, and lower load-shedding of $1-2\%$.

\begin{figure}
	\centering
	\includegraphics[width=.235\textwidth]{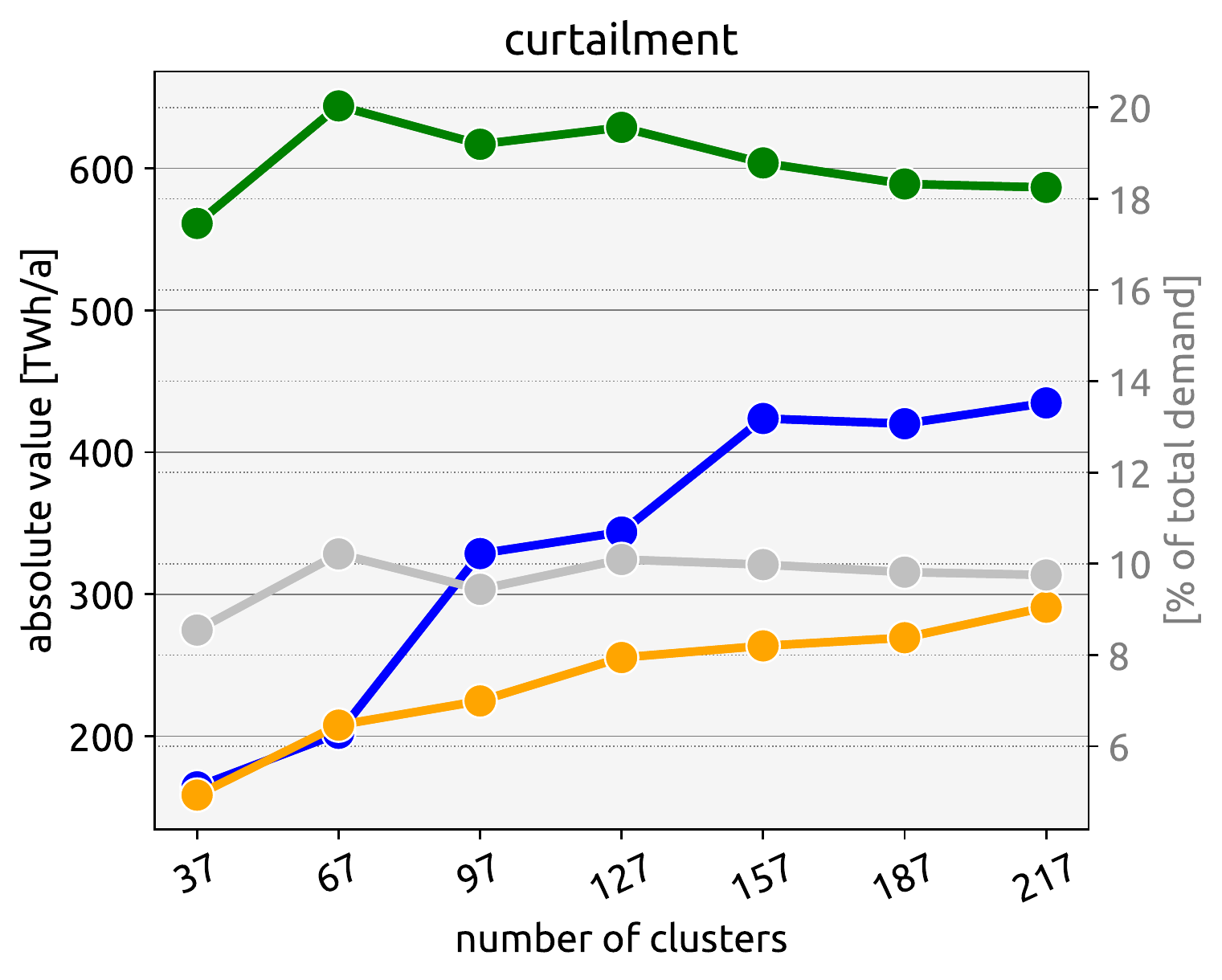}
	\includegraphics[width=.235\textwidth]{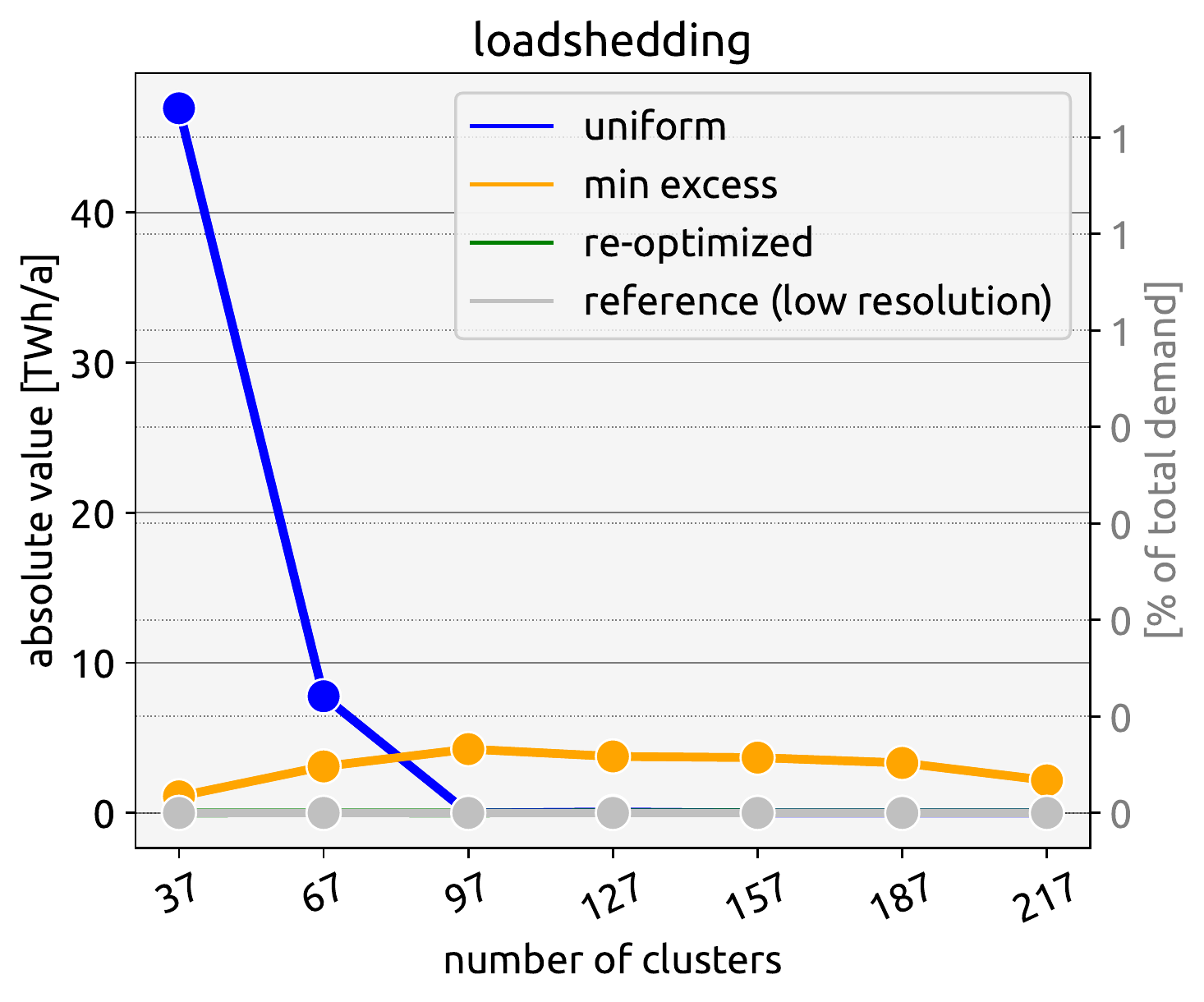}
	\caption{Results as displayed in Figure \ref{fig:loadsheddingandcurtailment_copperplate_island} of an island model, meaning that no inter-cluster electricity imports or exports are considered for the disaggregation.\label{fig:loadsheddingandcurtailment_copperplate_island}}
\end{figure}

\subsection{Analysing the Source for Load-Shedding} \label{sec:truefalse}

\begin{figure}
	\centering
	\begin{tabular}{c c}
		37 (reference) & 97 (reference) \\
		\includegraphics[trim=0 0 2.5cm 0, clip, width=.215\textwidth]{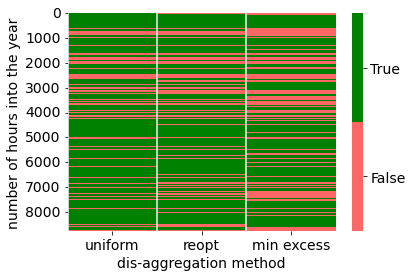} &
		\includegraphics[trim=0 0 0 0, clip,width=.26\textwidth]{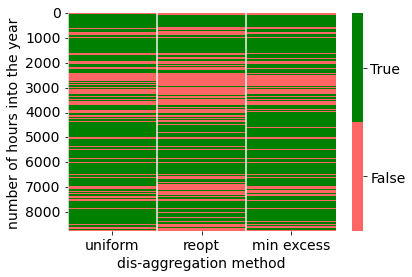} \hspace*{-.3cm} \\
		157 (reference) & 217 (reference) \\
		\includegraphics[trim=0 0 2.5cm 0, clip, width=.22\textwidth]{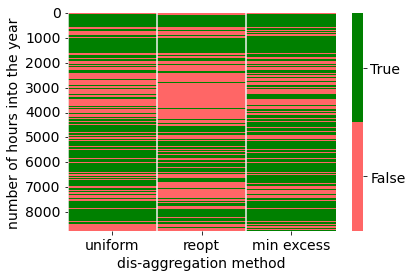} &
		\includegraphics[trim=0 0 2.5cm 0, clip, width=.22\textwidth]{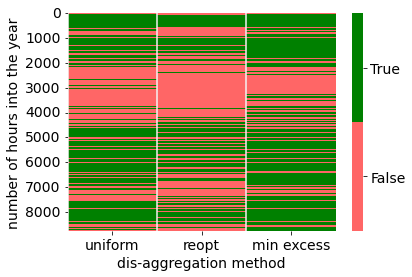} \hspace*{.52cm}
	\end{tabular}
	\caption{Evaluation if load-shedding measures of the dis-aggregated high-resolution models occur at times with higher curtailment compared to the lower resolved reference model, see equations (\ref{eq:truefalse1})-(\ref{eq:truefalse2}).} \label{fig:truefalse}
\end{figure}

Figure \ref{fig:truefalse} additionally displays if load-shedding measures occur at times where the curtailment of the high-resolution model is higher compared to the lower resolved reference results. If true, this indicates that the load-shedding measures are due to underestimated within-cluster transmission bottlenecks. To precisely evaluate this statement, the Figure displays the following hypothesis:
\begin{align} \label{eq:truefalse1}
	\delta_{\{t: (\Upsilon_{V}-\Upsilon_{C})(t) > 0\}} \cdot (\Upsilon_{V}-\Upsilon_{C})(t) &\geq 0.5 \cdot \Delta_{V}(t) \\
	\label{eq:truefalse2}
	\delta_{\{t:(\Upsilon_{V}-\Upsilon_{C})(t) < 0\}} \cdot (\Upsilon_{C} - \Upsilon_{V})(t) &\leq 0.5 \cdot \Delta_{V}(t) \,,
\end{align}
where $\Delta_{V}(t)$ represents the amount of load-shedding measures in the high-resolution dis-aggregated model at snapshot $t$, and $\Upsilon_{V}(t) := \sum_{\substack{s\in\mathcal{S}\\ v\in \mathcal{V}}}\left(\bar{g}_{v,s,t}G_{v,s}-g_{v,s,t}\right)$ the amount of curtailment in the high-resolution dis-aggregated model at snapshot $t$. Accordingly, $\Upsilon_{C}(t)$ represents the amount of curtailment in the lower resolved reference model.

It can be seen that, as the reference model resolution increases, there are more and more times $t$ where the hypothesis is wrong. The amount of curtailed electricity is higher than load-shedding in $85\%$ of the times on average for all of the three dis-aggregation methods for a very low-resolved reference model of $37$ nodes.
As the reference model resolution increases to $217$ nodes, the statement is only true in average for $65\%$ of the times for all three dis-aggregation methods. This indicates that transmission resolution is starting to saturate, however is still the major bottleneck preventing to feed-in the extra green electricity that is being curtailed.

\printcredits

\bibliographystyle{cas-model2-names}

\bibliography{inverse}

\bio{}
\endbio

\endbio

\end{document}